\documentclass[journal]{IEEEtran}
\usepackage{subfigure}
\usepackage{graphicx}
\usepackage{epsfig,latexsym,amsmath,graphics}
\usepackage{amssymb,dsfont,amsthm}
\usepackage{cite,color}
\usepackage{url}
\usepackage{booktabs}
\usepackage{algorithm,algorithmicx}
\usepackage{algpseudocode}
\usepackage{stfloats}
\usepackage{verbatim}
\usepackage{bm}
\usepackage{stfloats}
\usepackage{multirow}
\usepackage{ulem}
\usepackage{multirow}

\begin{document}
%	\pagewiselinenumbers % 按页重新编号 
%	\switchlinenumbers	 % 双栏，单栏删除该行
	\title{Weight Hybrid Architecture of Rydberg-Atomic Sensors}
	%\author{Hao~Wu,% <-this % stops a space
		%
		%}
	
	\author{Hao Wu, Xinyuan Yao, Shanchi Wu, Rui Ni, Chen Gong and Kaibin Huang
		% <-this % stops a space
		\thanks{This work was supported by National Natural Science Foundation of China	under Grant 62331024 and 62171428.}
		\thanks{Hao Wu, Xinyuan Yao, Shanchi Wu, Chen Gong are with the School of Information Science and Technology in University of Science and Technology of China, Email address: \{wuhao0719, yxy200127, wsc0807\}@mail.ustc.edu.cn, cgong821@ustc.edu.cn.
			
		Rui Ni is with Huawei Technology, Email address: raney.nirui@huawei.com.
		
		K. Huang is with the Department of Electronic and Electrical Engineering in the University of Hong Kong, Hong Kong, Email address: huangkb@eee.hku.hk.

}% <-this % stops a space
	}

\maketitle

\begin{abstract}
Rydberg atomic quantum receivers have been seen as novel radio frequency measurements and the high sensitivity to a large range of frequencies makes it attractive for communications reception. However, their performance can be significantly degraded by hardware-induced noise, particularly the noise from laser, which impacts the overall system noise floor and exhibits correlation. To address this challenge, this paper proposes a weight hybrid (WH) architecture for Rydberg-atomic sensors, a novel four-channel combining scheme designed for atomic sensors operating in correlated noise environments. By jointly processing dual signal channels and dual noise reference channels, the WH architecture effectively mitigates noise contributions from lasers and other hardware components. All channels are optimally combined via maximum likelihood estimation within an expectation maximization framework, enabling robust signal extraction under correlated noise. Moreover, the proposed WH architecture is universal and can be readily extended to other types of Rydberg receivers to achieve consistent performance improvements.

\end{abstract}

\section{Introduction}

Recently, Rydberg atoms have emerged as a novel platform for electric field sensing, enabling direct International System of Units (SI)-traceable and self-calibrated measurements\cite{anderson2021self,song2017quantum,holloway2014sub,simons2018electromagnetically,holloway2017electric}. The measured splitting only depends on the field strength, Planck’s constant, and dipole moment of the Rydberg transition, which can be reliably calculated. Rydberg atoms in highly excited states with one or more electrons of large principal quantum numbers are sensitive to electric fields, very suitable to manufacture atom-based sensor for detecting communication signals. Currently, this technique has been widely applied in diverse areas including polarization measurement \cite{sedlacek2013atom}, angle-of-arrival estimation \cite{robinson2021determining,mao2023digital}, subwavelength imaging \cite{holloway2017atom,downes2020full}, near-field antenna pattern characterization \cite{simons2019embedding,shi2023new}, and multi-frequency signal recognition \cite{liu2022deep,zhang2022rydberg,zhang2024image}.

﻿
From the perspective of Rydberg receiver architecture, improving the signal-to-noise ratio (SNR) fundamentally hinges on mitigating hardware-induced noise, particularly laser noise, which is a dominant performance-limiting factor \cite{jing2020atomic}. Two principal strategies have emerged in the development of this field: noise interference cancellation and intrinsically background-free detection methods. The multilevel structure of Rydberg atoms facilitates the deployment of both strategies by providing multiple accessible output channels that serve as either correlated noise references or background-free signal carriers.

\subsection{Related Works on Noise Interference Cancellation}
The first strategy employs noise interference cancellation, where a reference signal is used to subtract correlated hardware noise, as exemplified by balanced detection techniques in which a balanced avalanche photodiode (APD) simultaneously receives the signal-bearing probe laser (after field interaction) and a pure-noise reference probe laser (bypassing field interaction), effectively canceling common-mode laser noise to yield a low-noise signal, with alternative implementations such as Mach-Zehnder interferometers similarly relying on reference arms for noise suppression. As a classical optical signal processing architecture \cite{schlossberger2024rydberg,jing2020atomic,qimeng2025instantaneous,liang2026exceptional,jiang2025quantum,knarr2023spatiotemporal,manchaiah2026probing}, balanced detection leverages a reference probe signal for subtraction to eliminate the broad absorption dip inherent in probe scans while suppressing background laser power noise, though it requires meticulous balancing of the two APD inputs in practice. Despite this implementation challenge, as a fundamental and universally applicable processing method, it consistently enhances received signal quality across diverse applications—from achieving 55 nV/cm/Hz$^{1/2}$ sensitivity via superheterodyne architecture \cite{jing2020atomic} to integration with complex physical effects such as gradient magnetic fields for Rydberg electrometry \cite{qimeng2025instantaneous}, exceptional points in non-Hermitian systems \cite{liang2026exceptional}, quantum weak measurement for dispersion signal amplification \cite{jiang2025quantum}, and high-speed signal reception \cite{knarr2023spatiotemporal}, establishing balanced detection and its variants as a versatile processing paradigm that consistently improves received signal quality across a wide range of applications.
﻿ 
\subsection{Related Works on Background-free Detection}
The second, more fundamental strategy seeks to eliminate background noise at its source by using detection methods that are intrinsically background-free. Fluorescence detection \cite{teale2022degenerate,prajapati2024investigation} exemplifies this approach: although experimentally more complex and often requiring single-photon detectors, it bypasses the large coherent background inherent in electromagnetic induction transparency (EIT) readout, thereby enabling superior sensitivity with virtually no signal background. In addition, Works \cite{kumar2023quantum,borowka2024continuous} demonstrated the use of multiple energy levels to achieve microwave-to-optical conversion, where the absence of background laser noise enabled a sensitivity of 3.98  nV/cm/Hz$^{1/2}$\cite{borowka2024continuous}.
﻿

\subsection{Related Works on Multilevel Rydberg Structures}
Currently, two-color Rydberg-EIT has recently been extended to three-photon schemes \cite{thaicharoen2019electromagnetically,carr2012three,johnson2010absolute}. Work \cite{prajapati2023sensitivity} investigated the sensitivity of three-photon EIT for RF detection in Rydberg atoms by modeling four- and five-level systems to analyze probe transmission versus laser power and RF field strength, defining a shot-noise-limited sensitivity metric for comparison with the two-photon scheme. Due to the small dipole matrix elements involved in optical Rydberg excitation, conventional two-color implementations typically rely on expensive, high-performance coupling lasers. This limitation has motivated interest in alternative approaches employing three low-power infrared lasers. Beyond reducing laser cost and complexity, such three-photon extensions also enable all-infrared wavelength operation \cite{fahey2011excitation,johnson2012three,you2022microwave}, effectively suppress Doppler broadening, thereby facilitating narrower EIT linewidths \cite{bohaichuk2023three}, and provide a viable route for exciting high-angular-momentum Rydberg states, which are essential for electric field measurements in the very high frequency, ultra high frequency \cite{prajapati2024high,brown2023very}, and even terahertz regimes \cite{chen2022terahertz}.

\subsection{Motivation}

However, at present, regardless of whether it is the traditional ladder-type EIT architecture, more complex EIT schemes such as V-type configurations, or multi-laser mixing methods, signal readout invariably involves the detection of laser or fluorescence signals. A common limitation of these architectures is that they ultimately rely on only one or two channels, typically the probe laser or fluorescence signal for readout. Yet, in principle, all output channels, including the probe, coupling, dressing lasers, and fluorescence, carry partial information after interacting with the atomic vapor cell, even though some channels (e.g., coupling or dressing lasers) may exhibit poorer SNR compared to the probe laser. A straightforward generalization is therefore to jointly receive all available output channels from the physical system, rather than relying on a single laser output. Such a multi-channel receiving architecture would perform at least as well as the original single-channel approach, since the latter can be viewed as a special case where the probe channel weight is set to 1 and all others to 0. Furthermore, inspired by balanced detection techniques that use a reference laser (either bypassing the cell or passing through it without participating in the field interaction) to suppress noise and fluctuations, we can similarly incorporate these noise-only channels to optimally weight the signal-bearing channels. Unlike balanced detection, which relies on fixed analog weighting via balanced APDs, our approach enables adjustable weighting through digital signal processing, allowing for more flexible and optimized noise rejection.

In summary, these insights lead to a generalized architecture, the weight hybrid (WH) architecture, which utilizes all inputs to the Rydberg system as references for noise cancellation and all outputs as signal channels for joint estimation, thereby maximally suppressing hardware noise and achieving fundamentally enhanced detection performance.

\subsection{Contribution}

Our major contributions are summarized as follows:
\begin{itemize}
	\item[$\bullet$]The performance of Rydberg sensors is currently limited by hardware noise. Inspired by recent noise-mitigation strategies, we propose a WH architecture that jointly processes multi-channel outputs, moving beyond conventional single probe channel readout.
	
	\item[$\bullet$]For the case where the gain and noise characteristics of each channel are known, we derive the optimal weighting coefficients for the WH architecture and analyze the resulting SNR improvement over traditional single probe channel architectures under these ideal conditions.
	
	\item[$\bullet$]For the case where the gain and noise characteristics of each channel are unknown, we develop a weight and covariance matrix update algorithm for the WH architecture based on the MLE and an improved expectation maximization algorithm.
	
	\item[$\bullet$]We experimentally validated the superiority of the proposed WH architecture and discussed its generality and adaptability to more complex Rydberg sensing architectures.
\end{itemize}

\subsection{Organization}
The remainder of this paper is organized as follows. Section \ref{T3} presents and models the four sub-architectures of the WH framework, and analyzes their performance improvements over traditional architectures under the assumption of known covariance matrix and channel gains. In Section \ref{T4}, we derive the signal processing procedure for the WH architecture based on the maximum likelihood estimation (MLE) criterion and the expectation–maximization (EM) algorithm, addressing both the case of unknown covariance matrix and channel gains. Section \ref{T5} presents experimental results obtained from the proposed architecture. Section \ref{T6} provides an analysis of the scalability and extensibility of the WH architecture. Finally, Section \ref{T7} concludes this work.
\begin{figure*}
	\centering
	\includegraphics[width=0.89\textwidth]{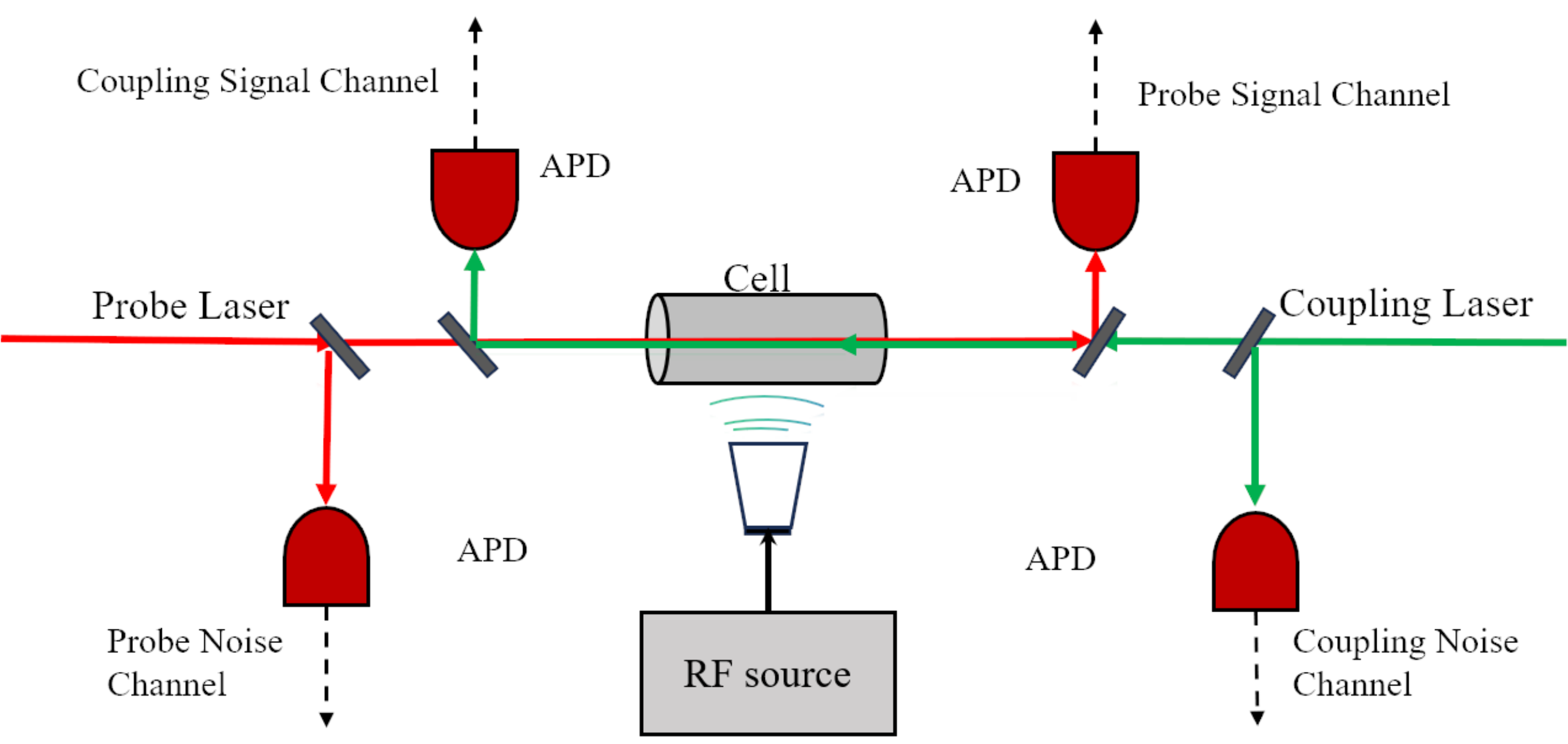}
	\caption{Schematic diagram of weight hybrid architecture in our work.}
	\label{diagram}
\end{figure*}

\begin{figure*}
	\centering
	\includegraphics[width=0.95\textwidth]{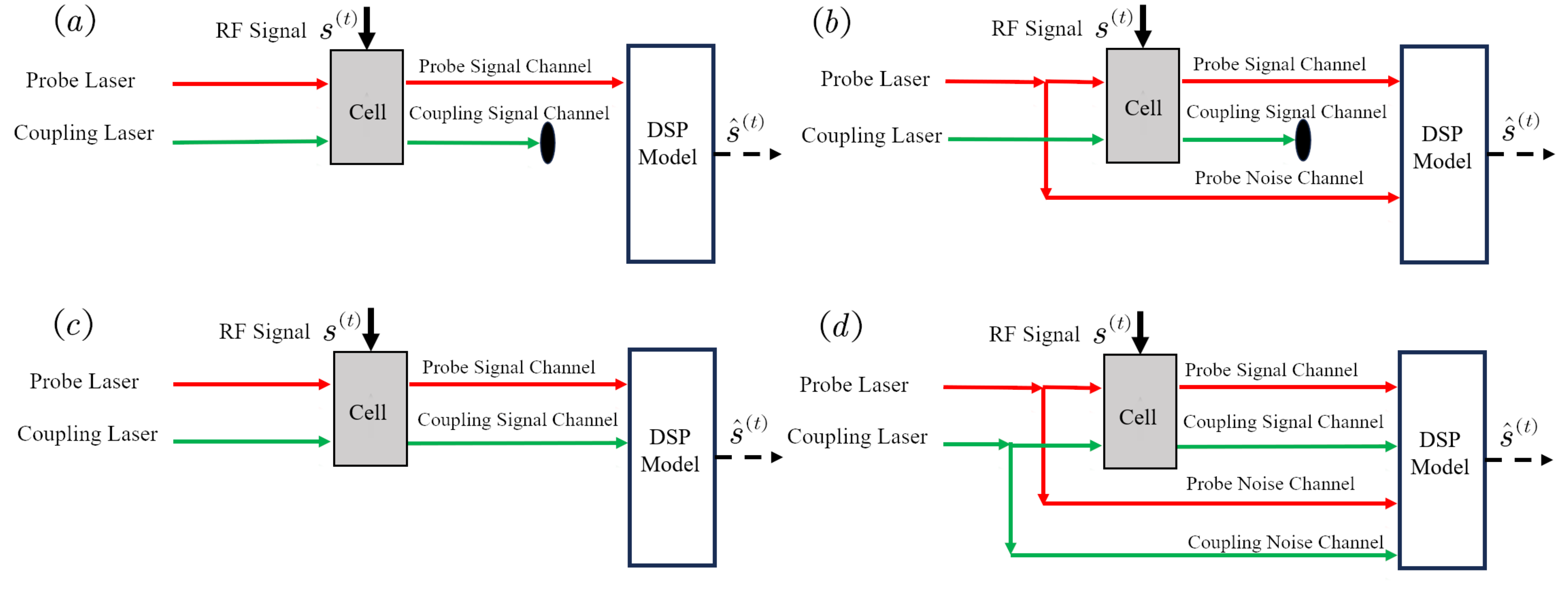}
	\caption{(a) WH-A architecture. (b) WH-B architecture. (c) WH-C architecture. (d) WH-D architecture}
	\label{Ediagram}
\end{figure*}

\section{WH Architecture with Unknown Gain and Noise}\label{T3}

\subsection{Model of Weight Hybrid Architecture}

Figure \ref{diagram} illustrates the weight hybrid architecture (WH) employed in our work, which comprises four channels of signal outputs (probe signal channel, coupling signal channel, probe noise channel, and coupling noise channel), rather than relying solely on the probe signal channel as in traditional Rydberg detection architectures. These four channels encompass all available information accessible in Rydberg atomic detection. The probe signal channel and coupling signal channel pass through the atomic vapor cell and participate in the four-level physical interaction, thus containing both signal and noise. The probe noise channel and coupling noise channel either do not pass through the atomic cell or pass through it without engaging in the four-level physical effect, thereby containing only noise.

Based on different utilization strategies of these four signal channels, Figure \ref{Ediagram} presents four hybrid weighting architectures. WH-A architecture shown in Figure \ref{Ediagram}(a) represents the current standard and commonly used Rydberg receiving architecture, which employs only the probe laser as the receiving signal detected by an APD, while the coupling laser is directed to an optical dump. WH-B architecture shown in Figure Figure \ref{Ediagram}(b) utilizes only the probe signal channel and the probe noise channel; compared to WH-A, it incorporates a beam splitter before the atomic cell to direct a portion of the probe laser to an additional APD, thereby requiring two APDs for probe detection (in a broad sense, WH-B architecture encompasses the balanced detection architecture adopted in some existing works). WH-C architecture shown in Figure \ref{Ediagram}(c) employs both the probe and coupling signal channels for reception; relative to WH-A, it adds reception of the coupling laser. WH-D architecture shown in Figure \ref{Ediagram}(d) encompasses reception of all available channels, including the probe signal channel, coupling signal channel, probe noise channel, and coupling noise channel. Obviously, due to the utilization of more information compared to the traditional standard architecture (WH-A), the WH-B, WH-C, and WH-D architectures, which employ more channels, can achieve better performance. Specifically, although the signal-to-noise ratio of the probe signal channel is superior to that of the coupling signal channel, the coupling signal channel also contains partial information. Meanwhile, although the noise channels contain no signal, the noise in these channels is correlated. Therefore, joint analysis integrating information from multiple channels can yield better results than single-channel reception.

The reception for the $t$-th symbol among $T$ symbols across the four channels can be expressed as
\begin{equation}
	\begin{aligned}
\boldsymbol{y}^{\left( t \right)}=\boldsymbol{h}s^{\left( t \right)}+\boldsymbol{n}^{\left( t \right)},
\end{aligned}
\end{equation}
where $s\left( t \right)\in \mathbb{C}$ and $y^{\left( t \right)}\in \mathbb{C}$ are the transmitted and received complex symbols, respectively. The gain of the $m$-th channel can be expressed in vector form, given by
\begin{equation}
	\begin{aligned}
\boldsymbol{h} =\left[ \begin{array}{c}
	h_1\\
	h_2\\
	0\\
	0\\
\end{array} \right] ,
\end{aligned}
\end{equation}
where $h_1\in \mathbb{C}$ and $h_2\in \mathbb{C}$ are the gains of the probe and coupling signal channels, respectively. $h_3=0$ and $h_4=0$ are the gains of the probe and coupling noise channels, respectively. The noise $\boldsymbol{n}$ follows $\mathbf{n}^{(t)} \sim \mathcal{CN}(\mathbf{0}, \mathbf{\Sigma})$ and
the covariance matrix $\boldsymbol{\Sigma }$ can be expressed as
\begin{equation}
	\begin{aligned}
\boldsymbol{\Sigma }&=\mathbb{E}\left[ \boldsymbol{nn}^H \right] 
\\
&=\left[ \begin{matrix}
	\sigma _{1}^{2}&		r_{12}\sigma _1\sigma _2&		r_{13}\sigma _1\sigma _3&		r_{14}\sigma _1\sigma _4\\
	r_{12}^{*}\sigma _1\sigma _2&		\sigma _{2}^{2}&		\rho _{23}\sigma _2\sigma _3&		r_{24}\sigma _2\sigma _4\\
	r_{13}^{*}\sigma _1\sigma _3&		r_{23}^{*}\sigma _2\sigma _3&		\sigma _{3}^{2}&		r_{34}\sigma _3\sigma _4\\
	r_{14}^{*}\sigma _1\sigma _4&		r_{24}^{*}\sigma _2\sigma _4&		r_{34}^{*}\sigma _3\sigma _4&		\sigma _{4}^{2}\\
\end{matrix} \right] ,
\end{aligned}
\end{equation}
where $\sigma_m^2 = \mathbb{E}[|n_m|^2]$ is the noise power of the $m$-th channel, and $r_{mk} = \frac{\mathbb{E}[n_m n_k]}{\sigma_m \sigma_k}$ is the complex correlation coefficient between channel $m$ and channel $k$, satisfying $|r_{mk}| \leq 1$. 

\subsection{Performance Analysis of Weight Hybrid Architecture}

We consider the case where the covariance matrix and channel gains are known (in practice, these quantities can be effectively measured experimentally). Then, the estimated symbol $\hat{s}^{\left( t \right)}$ for all four WH architecture based on the maximum likelihood criterion can be obtained as follows
\begin{equation}
	\begin{aligned}
\hat{s}^{\left( t \right)}&=\bm{w}^H\boldsymbol{y}^{\left( t \right)}
\\
&=\frac{\boldsymbol{h}^H\boldsymbol{\Sigma }^{-1}\boldsymbol{y}^{\left( t \right)}}{\boldsymbol{h}^H\boldsymbol{\Sigma }^{-1}\boldsymbol{h}},
\end{aligned}
\end{equation}
where $\bm{w}=\frac{\boldsymbol{\Sigma }^{-1}\boldsymbol{h}}{\boldsymbol{h}^H\boldsymbol{\Sigma }^{-1}\boldsymbol{h}}$ is the weight hybrid coefficient vector of the multi-channel WH architecture

Then, the estimation variance for each symbol, i.e., the noise variance under each WH architecture, can be expressed as
\begin{equation}
	\begin{aligned}
\mathbb{D}\left[ \hat{s}^{\left( t \right)} \right] &=\mathbb{D}\left[ \frac{\boldsymbol{h}^H\boldsymbol{\Sigma }^{-1}\left( \boldsymbol{h}s^{\left( t \right)}+\boldsymbol{n}^{\left( t \right)} \right)}{\boldsymbol{h}^H\boldsymbol{\Sigma }^{-1}\boldsymbol{h}} \right] 
\\
&=\mathbb{D}\left[ s^{\left( t \right)}+\frac{\boldsymbol{h}^H\boldsymbol{\Sigma }^{-1}\boldsymbol{n}^{\left( t \right)}}{\boldsymbol{h}^H\boldsymbol{\Sigma }^{-1}\boldsymbol{h}} \right] 
\\
&=\frac{1}{\boldsymbol{h}^H\boldsymbol{\Sigma }^{-1}\boldsymbol{h}}.
\end{aligned}
\end{equation}

\subsubsection{WH-A Architecture Model}
Since the WH-A architecture uses only one signal channel, the noise variance under the WH-A architecture can be expressed as
\begin{equation}
	\begin{aligned}
\mathbb{D}\left[ \hat{s}^{\left( t \right)} \right]_A &=\min_{m=1,2}\frac{\left[ \boldsymbol{\Sigma } \right] _{m,m}}{\left| h_m \right|^2},
\end{aligned}
\end{equation}
where $m=1$ denotes using only the probe signal channel, and $m=2$ denotes using only the coupling signal channel. Since the probe signal channel is generally superior to the coupling signal channel, let $m=1$, the noise variance under the WH-A architecture can typically be expressed as $\frac{\sigma_{1}^2}{\left| h_1 \right|^2}$, and the weight hybrid coefficient $\bm{w}_A=1/h_1$. Since the WH-A architecture is the traditional single probe channel receiver, its SNR gain relative to itself is 0 dB.

\subsubsection{WH-B Architecture Model}
Since the WH-B architecture uses the probe signal channel and the probe noise channel, we define $\boldsymbol{h}_{13}=\left[ h_1,h_3 \right] ^T=\left[ h_1,0 \right] ^T$. Then, the covariance matrix $\boldsymbol{\Sigma }_{13}$ for WH-B can be expressed as a submatrix of the four-channel WH-D covariance matrix $\boldsymbol{\Sigma }$, $\boldsymbol{\Sigma }_{13}$ and its inverse $\boldsymbol{\Sigma }_{13}^{-1}$ given by
\begin{equation}
	\begin{aligned}
		\boldsymbol{\Sigma }_{13}&=\left[ \begin{matrix}
			\sigma _{1}^{2}&		r_{13}\sigma _1\sigma _3\\
			r_{13}^{*}\sigma _1\sigma _3&		\sigma _{3}^{2}\\
		\end{matrix} \right] ,
		\\
		\boldsymbol{\Sigma }_{13}^{-1}&=\frac{1}{\sigma _{1}^{2}\sigma _{3}^{2}\left( 1-\left| r_{13} \right|^2 \right)}\left[ \begin{matrix}
			\sigma _{3}^{2}&		-r_{13}\sigma _1\sigma _3\\
			r_{13}^{*}\sigma _1\sigma _3&		\sigma _{1}^{2}\\
		\end{matrix} \right] .
	\end{aligned}
\end{equation}

Then, the noise variance under the WH-B architecture can be expressed as
\begin{equation}
	\begin{aligned}
		\mathbb{D}\left[ \hat{s}^{\left( t \right)} \right]_B &=\frac{1}{\boldsymbol{h}_{13}^{H}\boldsymbol{\Sigma }_{13}^{-1}\boldsymbol{h}_{13}}
		\\
		&=\frac{\sigma _{1}^{2}\left( 1-\left| r_{13} \right|^2 \right)}{\left| h_1 \right|^2}.
	\end{aligned}
\end{equation}

The SNR gain of WH-B architecture relative to the traditional single probe channel architecture (WH-A) can be expressed as
\begin{equation}
	\begin{aligned}
G_{BA}&=10\log _{10}\left( \frac{\text{SNR}_B}{\text{SNR}_A} \right) 
\\
&=10\log _{10}\left( \frac{\mathbb{D}\left[ \hat{s}^{\left( t \right)} \right] _A}{\mathbb{D}\left[ \hat{s}^{\left( t \right)} \right] _B} \right) 
\\
&=10\log _{10}\left( \frac{1}{1-\left| r_{13} \right|^2} \right) .
	\end{aligned}
\end{equation}

And the weight hybrid coefficient $\boldsymbol{w}_B$ of WH-B architecture is given by
\begin{equation}
	\begin{aligned}
\boldsymbol{w}_B&=\frac{\boldsymbol{\Sigma }_{13}^{-1}\boldsymbol{h}_{13}}{\boldsymbol{h}_{13}^{H}\boldsymbol{\Sigma }_{13}^{-1}\boldsymbol{h}_{13}}
\\
&=\frac{\mathbb{D}\left[ s^{\left( t \right)} \right] _B}{\sigma _{1}^{2}\sigma _{3}^{2}\left( 1-\left| r_{13} \right|^2 \right)}\left[ \begin{array}{c}
	\sigma _{3}^{2}h_1\\
	-r_{13}^{*}\sigma _1\sigma _3h_1\\
\end{array} \right] 
\\
&=\frac{1}{h_{1}^{*}}\left[ \begin{array}{c}
	1\\
	-r_{13}^{*}\frac{\sigma _1}{\sigma _3}\\
\end{array} \right] .
	\end{aligned}
\end{equation}

\subsubsection{WH-C Architecture Model}Since the WH-C architecture uses the probe signal channel and the coupling signal channel, we define $\boldsymbol{h}_{12}=\left[ h_1,h_2 \right] ^T$. Then, the covariance matrix $\boldsymbol{\Sigma }_{12}$ for WH-C can be expressed as a submatrix of the four-channel WH-D covariance matrix $\boldsymbol{\Sigma }$, $\boldsymbol{\Sigma }_{12}$ and its inverse $\boldsymbol{\Sigma }_{12}^{-1}$ given by
\begin{equation}
	\begin{aligned}
\boldsymbol{\Sigma }_{12}&=\left[ \begin{matrix}
	\sigma _{1}^{2}&		r_{12}\sigma _1\sigma _2\\
	r_{12}^{*}\sigma _1\sigma _2&		\sigma _{2}^{2}\\
\end{matrix} \right] ,
\\
\boldsymbol{\Sigma }_{12}^{-1}&=\frac{1}{\sigma _{1}^{2}\sigma _{2}^{2}\left( 1-\left| r_{12} \right|^2 \right)}\left[ \begin{matrix}
	\sigma _{2}^{2}&		-r_{12}\sigma _1\sigma _2\\
	r_{12}^{*}\sigma _1\sigma _2&		\sigma _{1}^{2}\\
\end{matrix} \right] .
\end{aligned}
\end{equation}

Then, the noise variance under the WH-C architecture can be expressed as
\begin{equation}
	\begin{aligned}
\mathbb{D}\left[ \hat{s}^{\left( t \right)} \right]_C &=\frac{1}{\boldsymbol{h}_{12}^{H}\boldsymbol{\Sigma }_{12}^{-1}\boldsymbol{h}_{12}}
\\
&=\frac{\sigma_{1}^2\sigma_{2}^2\left( 1-\left|r_{12} \right|^2 \right) }{\left| h_1 \right|^2\sigma _{2}^{2}+\left| h_2 \right|^2\sigma _{1}^{2}-2\Re \left[ r_{12}h_{1}^{*}h_2\sigma _1\sigma _2 \right]}.
\end{aligned}
\end{equation}

The SNR gain of WH-C architecture relative to the traditional single probe channel architecture (WH-A) can be expressed as
\begin{equation}
	\begin{aligned}
		G_{CA}&=10\log _{10}\left( \frac{\mathbb{D}\left[ \hat{s}^{\left( t \right)} \right] _A}{\mathbb{D}\left[ \hat{s}^{\left( t \right)} \right] _C} \right) 
		\\
		&=10\log _{10}\left( \frac{\left| h_1 \right|^2\sigma _{2}^{2}+\left| h_2 \right|^2\sigma _{1}^{2}-2\sigma _1\sigma _2\Re \left[ r_{12}h_{1}^{*}h_2 \right]}{\left| h_1 \right|^2\sigma _{2}^{2}\left( 1-\left| r_{12} \right|^2 \right)} \right) .
	\end{aligned}
\end{equation}

And the weight hybrid coefficient $\boldsymbol{w}_C$ of WH-C architecture is given by
\begin{equation}
	\begin{aligned}
\boldsymbol{w}_C&=\frac{\boldsymbol{\Sigma }_{12}^{-1}\boldsymbol{h}_{12}}{\boldsymbol{h}_{12}^{H}\boldsymbol{\Sigma }_{12}^{-1}\boldsymbol{h}_{12}}
\\
&=\frac{\mathbb{D}\left[ s^{\left( t \right)} \right] _C}{\sigma _{1}^{2}\sigma _{2}^{2}\left( 1-\left| r_{12} \right|^2 \right)}\left[ \begin{array}{c}
	\sigma _{2}^{2}h_1-r_{12}\sigma _1\sigma _2h_2\\
	-r_{12}^{*}\sigma _1\sigma _2h_1+\sigma _{1}^{2}h_2\\
\end{array} \right] .
\end{aligned}
\end{equation}

\subsubsection{WH-D Architecture Model}
The WH-D architecture utilizes information from all four channels. We define $\bm{h} = [\bm{h}_s; \mathbf{0}]$ and $ \bm{h}_s = [h_1 , h_2]^T $, so the covariance matrix $\bm{\Sigma}$ and its inverse $\bm{\Sigma}^{-1}$ can be expressed as

\begin{equation}
	\begin{aligned}
\boldsymbol{\Sigma }&=\left[ \begin{matrix}
	\boldsymbol{\Sigma }_{ss}&		\boldsymbol{\Sigma }_{sn}\\
	\boldsymbol{\Sigma }_{sn}^{H}&		\boldsymbol{\Sigma }_{nn}\\
\end{matrix} \right] ,
\\
\boldsymbol{\Sigma }^{-1}&=\left[ \begin{matrix}
	\boldsymbol{A}&		\boldsymbol{B}\\
	\boldsymbol{B}^H&		\boldsymbol{C}\\
\end{matrix} \right] ,
\end{aligned}
\end{equation}
where $\boldsymbol{\Sigma}_{ss} = \boldsymbol{\Sigma}_{12}$ is the $2\times2$ covariance matrix of the signal channels, $ \boldsymbol{\Sigma}_{nn} $ is the $2\times2$ covariance matrix of the noise channels, and $ \boldsymbol{\Sigma}_{sn} $ is the $2\times2$ covariance matrix between the signal and noise channels. Based on block matrix inversion, the upper-left $ 2 \times 2 $ submatrix $ \bm{A} $ of the inverse covariance matrix $\bm{\Sigma}^{-1}$ can be expressed as $\boldsymbol{A}=\left( \boldsymbol{\Sigma }_{ss}-\boldsymbol{\Sigma }_{sn}\boldsymbol{\Sigma }_{nn}\boldsymbol{\Sigma }_{sn}^{H} \right) ^{-1}$, and the upper-right submatrix $\boldsymbol{B}=-\boldsymbol{A\Sigma }_{sn}\boldsymbol{\Sigma }_{nn}^{-1}$.

Then, the noise variance under the WH-D architecture can be expressed as
\begin{equation}
	\begin{aligned}
\mathbb{D}\left[ \hat{s}^{\left( t \right)} \right]_D &=\frac{1}{\left[ \boldsymbol{h}_s;0 \right] ^H\boldsymbol{\Sigma }^{-1}\left[ \boldsymbol{h}_s;0 \right]}
\\
&=\frac{1}{\boldsymbol{h}_{s}^{H}\boldsymbol{Ah}_s}
\\
&=\frac{1}{\boldsymbol{h}_{s}^{H}\left( \boldsymbol{\Sigma }_{ss}-\boldsymbol{\Sigma }_{sn}\boldsymbol{\Sigma }_{nn}\boldsymbol{\Sigma }_{sn}^{H} \right) ^{-1}\boldsymbol{h}_s}.
\end{aligned}
\end{equation}

The SNR gain of WH-D architecture relative to the traditional single probe channel architecture (WH-A) can be expressed as
\begin{equation}
	\begin{aligned}
		G_{DA}&=10\log _{10}\left( \frac{\mathbb{D}\left[ \hat{s}^{\left( t \right)} \right] _A}{\mathbb{D}\left[ \hat{s}^{\left( t \right)} \right] _D} \right) 
		\\
		&=10\log _{10}\left( \frac{\sigma _{1}^{2}}{\left| h_1 \right|^2}\boldsymbol{h}_{s}^{H}\boldsymbol{Ah}_s \right) .
	\end{aligned}
\end{equation}

And the weight hybrid coefficient $\boldsymbol{w}_D$ of WH-D architecture is given by
\begin{equation}
	\begin{aligned}
\boldsymbol{w}_D&=\frac{\boldsymbol{\Sigma }^{-1}\boldsymbol{h}}{\boldsymbol{h}^H\boldsymbol{\Sigma }^{-1}\boldsymbol{h}}
\\
&=\mathbb{D}\left[ s^{\left( t \right)} \right] _D\left[ \begin{array}{c}
	\boldsymbol{Ah}_s\\
	\boldsymbol{B}^H\boldsymbol{h}_s\\
\end{array} \right] 
\\
&=\frac{1}{\boldsymbol{h}_{s}^{H}\boldsymbol{Ah}_s}\left[ \begin{array}{c}
	\boldsymbol{Ah}_s\\
	\boldsymbol{B}^H\boldsymbol{h}_s\\
\end{array} \right] .
\end{aligned}
\end{equation}

\subsection{Performance Comparison of Weight Hybrid Architecture}

\subsubsection{Performance Comparison of WH Architecture with Different Numbers of Channels}
Intuitively, WH architectures that utilize more channels can achieve better performance. Mathematically, since the covariance matrix $ \boldsymbol{\Sigma}_{ss} $ (i.e., $ \boldsymbol{\Sigma}_{12}$) is positive definite, and the matrix $ \boldsymbol{\Sigma}_{sn} \boldsymbol{\Sigma}_{nn}^{-1} \boldsymbol{\Sigma}_{sn}^{H}$ is positive semidefinite, we have $\boldsymbol{\Sigma}_{ss} - \boldsymbol{\Sigma}_{sn} \boldsymbol{\Sigma}_{nn}^{-1} \boldsymbol{\Sigma}_{sn}^{H} \preceq \boldsymbol{\Sigma}_{ss}$. Taking the inverse of both sides (which reverses the matrix inequality due to the monotonicity of the inverse for positive definite matrices) yields $\left( \boldsymbol{\Sigma}_{ss} - \boldsymbol{\Sigma}_{sn} \boldsymbol{\Sigma}_{nn}^{-1} \boldsymbol{\Sigma}_{sn}^{H} \right)^{-1} \succeq \boldsymbol{\Sigma}_{ss}^{-1}$. Thus, for any given signal channel gain vector $ \boldsymbol{h}_s $, we obtain $\boldsymbol{h}_s^{H} \left( \boldsymbol{\Sigma}_{ss} - \boldsymbol{\Sigma}_{sn} \boldsymbol{\Sigma}_{nn}^{-1} \boldsymbol{\Sigma}_{sn}^{H} \right)^{-1} \boldsymbol{h}_s \geq \boldsymbol{h}_s^{H} \boldsymbol{\Sigma}_{ss}^{-1} \boldsymbol{h}_s$. Additionally, for a positive definite matrix $ \boldsymbol{\Sigma}_{ss} $, the following lower bound relationship exists between its diagonal elements and the vector inner product $\boldsymbol{h}_s^{H} \boldsymbol{\Sigma}_{ss}^{-1} \boldsymbol{h}_s \geq \frac{\left| [\boldsymbol{h}_s]_m \right|^2}{[\boldsymbol{\Sigma}_{ss}]_{m,m}}$. Combining the above relationships, we finally obtain the inequality chain for the estimation variance under each architecture
\begin{equation}
	\begin{aligned}
\frac{1}{\boldsymbol{h}_s^{H} \left( \boldsymbol{\Sigma}_{ss} - \boldsymbol{\Sigma}_{sn} \boldsymbol{\Sigma}_{nn}^{-1} \boldsymbol{\Sigma}_{sn}^{H} \right)^{-1} \boldsymbol{h}_s} \leq \frac{1}{\boldsymbol{h}_s^{H} \boldsymbol{\Sigma}_{ss}^{-1} \boldsymbol{h}_s} \leq \frac{[\boldsymbol{\Sigma}_{ss}]_{m,m}}{\left| [\boldsymbol{h}_s]_m \right|^2}.
\end{aligned}
\end{equation}

Based on the above analysis, the estimation variances of the four WH architectures satisfy the following relationship
\begin{equation}
	\begin{aligned}
\mathbb{D}\left[ s^{\left( t \right)} \right] _D&\leq \mathbb{D}\left[ s^{\left( t \right)} \right] _B\leq \mathbb{D}\left[ s^{\left( t \right)} \right] _A,
\\
\mathbb{D}\left[ s^{\left( t \right)} \right] _D&\leq \mathbb{D}\left[ s^{\left( t \right)} \right] _C\leq \mathbb{D}\left[ s^{\left( t \right)} \right] _A.
\end{aligned}
\end{equation}

\subsubsection{Performance Comparison of WH Architecture with Different Numbers of Channels}

\subsubsection{Performance Comparison of WH-B Architecture with WH-A Architecture}
The advantage of the WH-B architecture over the WH-A architecture stems from the noise reference. The probe noise channel provides a noise observation correlated with the probe signal channel, mathematically expressed as $r_{13}$. Since the probe noise channel originates from the same probe laser and shares identical laser and noise sources, it exhibits high correlation $r_{13}$, thereby improving performance. The advantage of the WH-B architecture over balanced detection lies in its use of digital, noise-information-based adjustable hybrid weighting to combine information from the probe signal channel and the noise channel, rather than simply employing analog, fixed-weight balanced detection APD to weight information from two channels. This allows for practical deployment using two conventional APDs instead of a balanced APD, while the digital weighting mechanism eliminates the need for complex adjustments of the incident laser power to the balanced APD during implementation. However, this comes at the cost of increased hardware complexity, WH-B requires extra APD and ADC compared to single-APD receivers.

\subsubsection{Performance Comparison of WH-C Architecture with WH-A Architecture}
The advantage of the WH-C architecture over the WH-A architecture stems from additional signal power $h_2$ and noise reference $r_{12}$. Although the coupling signal channel is weaker, it still contains signal information, thereby providing additional signal power. Furthermore, the coupling signal channel offers a noise observation correlated with the probe signal channel, mathematically expressed as $r_{12}$. Although the probe and coupling signal channels originate from separate probe and coupling lasers, the EIT frequency-locking system establishes a correlation between the two lasers. Moreover, the interaction of the two laser beams within the atomic cell further induces noise correlation between the probe and coupling signal channels, thereby enhancing performance. And WH-C requires extra APD and ADC compared to single-APD receivers.

\subsubsection{Performance Comparison of WH-B Architecture with WH-C Architecture}

The performance advantage between WH-B and WH-C architectures is evaluated using the following inequality
\begin{equation}
	\begin{aligned}
\frac{1}{1+\frac{\left| h_2 \right|^2\sigma _{1}^{2}-2\sigma _1\sigma _2\Re \left[ r_{12}h_{1}^{*}h_2 \right]}{\left| h_1 \right|^2\sigma _{2}^{2}}}\left( 1-\left| r_{12} \right|^2 \right) \gtrless \left( 1-\left| r_{13} \right|^2 \right) ,
\end{aligned}
\end{equation}
where which architecture performs better depends on specific device characteristics, including channel gains, noise variances, and correlation coefficients, so the optimal choice is determined by the relative strength of the coupling channel and the achievable noise correlation in a given experimental setup.

\section{WH Architecture with Unknown Gain and Noise}\label{T4}

For the WH-A and WH-C architectures, all channels contain signal components, so the channel gains $\bm{h}$ are non-zero. The difference lies in the number of channels: WH-A uses a single signal channel, while WH-C uses two signal channels. For the WH-B and WH-D architectures, the channel gain vector $\bm{h}$ contains both zero and non-zero components, corresponding to noise-only channels and signal-bearing channels, respectively. The difference lies in the number of channels: WH-B uses one signal channel and one noise channel, while WH-D uses two signal channels and two noise channels.

\subsection{Expectation-Maximization Algorithm}\label{T41}
Without loss of generality, we a total of $N_s$ signal channels and $N_n$ noise channels, resulting in $N=N_s+N_n$ reception channels. For the $t$-th transmitted symbol $s^{\left(t \right) }$, the observed vector from all $N$ channels is denoted as 
 \begin{equation}
 	\begin{aligned}
\boldsymbol{y}^{\left( t \right)}=\left[ \begin{array}{c}
	\boldsymbol{y}_{s}^{\left( t \right)}\\
	\boldsymbol{y}_{n}^{\left( t \right)}\\
\end{array} \right] ,
	\end{aligned}
\end{equation}
where $\boldsymbol{y}_{s}^{\left( t \right)} \in \mathbb{C}^{N_s\times 1}$ correspond to the signal channels and $\boldsymbol{y}_{n}^{\left( t \right)} \in \mathbb{C}^{N_n\times 1}$ correspond to the noise channels. 

The transmitted symbol $s^{\left( t \right)}$ is an M-QAM symbol taken from a finite alphabet set $\mathcal{A} = \left\lbrace a_1, a_2, ..., a_M\right\rbrace $. For the $t$-th transmitted symbol $s^{\left(t \right) }=a_m$, the conditional probability density function of $\boldsymbol{y}^{\left( t \right)}$ is given by
\begin{equation}
	\begin{aligned}
		&\mathbb{P}\left[ \boldsymbol{y}^{\left( t \right)}|s^{\left( t \right)}=a_m,\boldsymbol{\theta } \right] =\frac{1}{\pi ^{N}\left| \boldsymbol{\Sigma } \right|}
		\\
		& \ \ \ \  \ \ \ \  \times \exp \left[ -\left( \boldsymbol{y}^{\left( t \right)}-\boldsymbol{h}a_m \right) ^H\boldsymbol{\Sigma }^{-1}\left( \boldsymbol{y}^{\left( t \right)}-\boldsymbol{h}a_m \right) \right] ,
	\end{aligned}
\end{equation}
where the parameter set is denoted as $\boldsymbol{\theta} = \{\boldsymbol{h}_s, \boldsymbol{\Sigma}\}$. The channel gain vector is
\begin{equation}
	\begin{aligned}
\boldsymbol{h}=\left[ \begin{array}{c}
	\boldsymbol{h}_s\\
	\boldsymbol{h}_n\\
\end{array} \right],
	\end{aligned}
\end{equation}
where $\boldsymbol{h}_s\in \mathbb{C}^{N_s\times 1}$ correspond to the signal channels and the remaining $N_n$ entries $\boldsymbol{h}_n=\bm{0}$ to the noise channels (where noise channel gains are zero). The noise covariance matrix $\boldsymbol{\Sigma} \in \mathbb{C}^{N \times N}$ characterizes the correlation among noise components across all channels. According to the structure of $\boldsymbol{h}$, the covariance matrix $\boldsymbol{\Sigma}$ is partitioned as
\begin{equation}\label{Sigma_Block}
	\begin{aligned}
		\boldsymbol{\Sigma }&=\left[ \begin{matrix}
			\boldsymbol{\Sigma }_{ss}&		\boldsymbol{\Sigma }_{sn}\\
			\boldsymbol{\Sigma }_{sn}^{H}&		\boldsymbol{\Sigma }_{nn}\\
		\end{matrix} \right] ,
		\\
		\boldsymbol{\Sigma }^{-1}&=\left[ \begin{matrix}
			\boldsymbol{A}&		\boldsymbol{B}\\
			\boldsymbol{B}^H&		\boldsymbol{C}\\
		\end{matrix} \right] ,
	\end{aligned}
\end{equation}
where $\boldsymbol{\Sigma}_{ss}$ is the $N_s\times N_s$ covariance matrix of the signal channels, $ \boldsymbol{\Sigma}_{nn} $ is the $N_n \times N_n $ covariance matrix of the noise channels, and $ \boldsymbol{\Sigma}_{sn} $ is the $N_s\times N_n$ covariance matrix between the signal and noise channels. Based on block matrix inversion, $\boldsymbol{A}=\left( \boldsymbol{\Sigma }_{ss}-\boldsymbol{\Sigma }_{sn}\boldsymbol{\Sigma }_{nn}\boldsymbol{\Sigma }_{sn}^{H} \right) ^{-1}$,  $\boldsymbol{B}=-\boldsymbol{A\Sigma }_{sn}\boldsymbol{\Sigma }_{nn}^{-1}$ , and $\boldsymbol{C}=\boldsymbol{\Sigma }_{nn}^{-1}+\boldsymbol{\Sigma }_{nn}^{-1}\boldsymbol{\Sigma }_{sn}^{H}\boldsymbol{A\Sigma }_{sn}\boldsymbol{\Sigma }_{nn}^{-1}$. When $N_n=0$, we have $ \boldsymbol{\Sigma}_{sn}=\boldsymbol{0}$, $ \boldsymbol{\Sigma}_{nn}=\boldsymbol{0}$, $\boldsymbol{A}= \boldsymbol{\Sigma }_{ss}^{-1}$,  $\boldsymbol{B}=\boldsymbol{0}$ , and $\boldsymbol{C}=\boldsymbol{0}$.

For all $T$ observations, the joint log-likelihood of the complete data $\boldsymbol{Y}=\left\{ \boldsymbol{y}^{\left( 1 \right)},\cdots ,\boldsymbol{y}^{\left( t \right)},\cdots ,\boldsymbol{y}^{\left( T \right)} \right\} 
$ and $\boldsymbol{S}=\left\{ s^{\left( 1 \right)},\cdots ,s^{\left( t \right)},\cdots ,s^{\left( T \right)} \right\} $ is given by:
\begin{equation}
	\begin{aligned}
		&\ln \mathbb{P}\left[ \boldsymbol{Y,S|\theta } \right] =\sum_{t=1}^T{\ln \mathbb{P}\left[ \boldsymbol{y}^{\left( t \right)},s^{\left( t \right)}|\boldsymbol{\theta } \right]}+\sum_{t=1}^T{\ln \mathbb{P}\left[ s^{\left( t \right)} \right]}
		\\
		&=-\sum_{t=1}^T{\left( \boldsymbol{y}^{\left( t \right)}-\boldsymbol{h}s^{\left( t \right)} \right) ^H\boldsymbol{\Sigma }^{-1}\left( \boldsymbol{y}^{\left( t \right)}-\boldsymbol{h}s^{\left( t \right)} \right)}
		\\
		&\ \ \ \ \ \ -T\ln \left| \boldsymbol{\Sigma } \right|+C,
	\end{aligned}
\end{equation}
where $C = -NT\ln\pi - T\ln M$ is a constant arising from the uniform prior distribution of the transmitted symbols $s^{\left( t\right) }$. The EM algorithm iterates between an expectation step and a maximization step to refine $\boldsymbol{\theta }$.

\subsubsection{Expectation Step}

In the E-step, we compute the posterior probability $w_{m}^{\left( t \right)}=\mathbb{P}\left[ s^{\left( t \right)}=a_m|\boldsymbol{y}^{\left( t \right)},\boldsymbol{\theta}^{\left[ k\right] } \right]$ of each symbol given the current parameter estimates $\boldsymbol{\theta}^{\left[ k\right] } = \{\boldsymbol{h}_s^{\left[ k\right] }, \boldsymbol{\Sigma}^{\left[ k\right] }\} $ and the observation $ \boldsymbol{y}^{(t)}$ , which is then used to form the Q-function $Q\left( \boldsymbol{\theta |\theta }^{\left[ k \right]} \right) =\mathbb{E}_{\boldsymbol{S|Y,\theta }^{\left[ k \right]}}\left[ \ln \mathbb{P}\left[ \boldsymbol{Y,S|\theta } \right] \right] $, given by
\begin{equation}
	\begin{aligned}
		&w_{m}^{\left( t \right)}=\mathbb{P}\left[ s^{\left( t \right)}=a_m|\boldsymbol{y}^{\left( t \right)},\boldsymbol{\theta }^{\left( k \right)} \right] 
		\\
		&=\frac{\mathbb{P}\left[ \boldsymbol{y}^{\left( t \right)}|s^{\left( t \right)}=a_m,\boldsymbol{\theta }^{\left( k \right)} \right] \mathbb{P}\left[ s^{\left( t \right)}=a_m \right]}{\sum_{j=1}^M{\mathbb{P}\left[ \boldsymbol{y}^{\left( t \right)}|s^{\left( t \right)}=a_j,\boldsymbol{\theta }^{\left( k \right)} \right] \mathbb{P}\left[ s^{\left( t \right)}=a_j \right]}}
		\\
		&=\frac{\exp \left[ -\left( \boldsymbol{y}^{\left( t \right)}-\boldsymbol{h}^{\left[ k \right]}a_m \right) ^H\left( \boldsymbol{\Sigma }^{\left[ k \right]} \right) ^{-1}\left( \boldsymbol{y}^{\left( t \right)}-\boldsymbol{h}^{\left[ k \right]}a_m \right) \right]}{\sum_{j=1}^M{\exp \left[ -\left( \boldsymbol{y}^{\left( t \right)}-\boldsymbol{h}^{\left[ k \right]}a_j \right) ^H\left( \boldsymbol{\Sigma }^{\left[ k \right]} \right) ^{-1}\left( \boldsymbol{y}^{\left( t \right)}-\boldsymbol{h}^{\left[ k \right]}a_j \right) \right]}}
		\\
		&=\frac{\exp \left[ -d_{m}^{\left( t \right)} \right]}{\sum_{j=1}^M{\exp \left[ -d_{j}^{\left( t \right)} \right]}},
	\end{aligned}
\end{equation}
where we assume equally likely symbols, i.e., $\mathbb{P}\left[ s^{\left( t \right)}=a_m \right] =1/M$. Using the partitioned structure of $\left( \boldsymbol{\Sigma }^{\left[ k \right]}\right) ^{-1}$, $d_{m}^{\left( t \right)}$ can be expressed as
\begin{equation}\label{d_equation}
	\begin{aligned}
		d_{m}^{\left( t \right)}=&\left( \boldsymbol{y}^{\left( t \right)}-\boldsymbol{h}^{\left[ k \right]}a_m \right) ^H\left( \boldsymbol{\Sigma }^{\left[ k \right]} \right) ^{-1}\left( \boldsymbol{y}^{\left( t \right)}-\boldsymbol{h}^{\left[ k \right]}a_m \right)
		\\
		=&\left( \boldsymbol{y}_{s}^{\left( t \right)}-\boldsymbol{h}_{s}^{\left[ k \right]}a_m \right) ^H\boldsymbol{A}^{\left[ k \right]}\left( \boldsymbol{y}_{s}^{\left( t \right)}-\boldsymbol{h}_{s}^{\left[ k \right]}a_m \right) 
		\\
		&+\left( \boldsymbol{y}_{n}^{\left( t \right)} \right) ^H\boldsymbol{C}^{\left[ k \right]}\boldsymbol{y}_{n}^{\left( t \right)}+\left( \boldsymbol{y}_{s}^{\left( t \right)}-\boldsymbol{h}_{s}^{\left[ k \right]}a_m \right) ^H\boldsymbol{B}^{\left[ k \right]}\boldsymbol{y}_{n}^{\left( t \right)}
		\\
		&+\left( \boldsymbol{y}_{n}^{\left( t \right)} \right) ^H\left( \boldsymbol{B}^{\left[ k \right]} \right) ^H\left( \boldsymbol{y}_{s}^{\left( t \right)}-\boldsymbol{h}_{s}^{\left[ k \right]}a_m \right) ,
	\end{aligned}
\end{equation}
where $\boldsymbol{A}^{\left[ k \right]}$, $\boldsymbol{B}^{\left[ k \right]}$ and $\boldsymbol{C}^{\left[ k \right]}$ can be expressed as

\begin{equation}\label{ABC_equation}
	\begin{aligned}
\boldsymbol{A}^{\left[ k \right]}&=\left( \boldsymbol{\Sigma }_{ss}^{\left[ k \right]}-\boldsymbol{\Sigma }_{sn}^{\left[ k \right]}\left( \boldsymbol{\Sigma }_{nn}^{\left[ k \right]} \right) ^{-1}\left( \boldsymbol{\Sigma }_{sn}^{\left[ k \right]} \right) ^H \right) ^{-1},
\\
\boldsymbol{B}^{\left[ k \right]}&=-\boldsymbol{A}^{\left[ k+1 \right]}\boldsymbol{\Sigma }_{sn}^{\left[ k \right]}\left( \boldsymbol{\Sigma }_{nn}^{\left[ k\right]} \right) ^{-1},
\\
\boldsymbol{C}^{\left[ k \right]}&=\left( \boldsymbol{\Sigma }_{nn}^{\left[ k \right]} \right) ^{-1}+\left( \boldsymbol{\Sigma }_{nn}^{\left[ k\right]} \right) ^{-1}\left( \boldsymbol{\Sigma }_{sn}^{\left[ k \right]} \right) ^H\boldsymbol{A}^{\left[ k+1 \right]}\boldsymbol{\Sigma }_{sn}^{\left[ k \right]}\left( \boldsymbol{\Sigma }_{nn}^{\left[ k \right]} \right) ^{-1},
	\end{aligned}
\end{equation}
where $\boldsymbol{A}^{\left[ k \right]}=\left( \boldsymbol{\Sigma }_{ss}^{\left[ k \right]} \right) ^{-1}$, $\boldsymbol{B}^{\left[ k \right]}=\boldsymbol{0}$ and $\boldsymbol{C}^{\left[ k \right]}=\boldsymbol{0}$ for $N_n =0$.

For numerical stability, we define $d_{\max}=max\ \left\{ d_{1}^{\left( t \right)},d_{2}^{\left( t \right)},\cdots ,d_{M}^{\left( t \right)} \right\} $ and $w_{m}^{\left( t \right)}$ is given by
\begin{equation}\label{w_equation}
	\begin{aligned}
		w_{m}^{\left( t \right)}=\frac{\exp \left[ -\left( d_{m}^{\left( t \right)}-d_{\max} \right) \right]}{\sum_{j=1}^M{\exp \left[ -\left( d_{m}^{\left( t \right)}-d_{\max} \right) \right]}}.
	\end{aligned}
\end{equation}

Based on these posterior probabilities, we can derive the conditional first and second moments of the transmitted symbols, as well as their conditional variance:
\begin{equation}\label{m_equation}
	\begin{aligned}
		\mathbb{E}\left[ s^{\left( t \right)}|\boldsymbol{y}^{\left( t \right)},\boldsymbol{\theta }^{\left[ k \right]} \right] &=\sum_{m=1}^M{a_mw_{m}^{\left( t \right)}},
		\\
		\mathbb{E}\left[ \left| s^{\left( t \right)} \right|^2|\boldsymbol{y}^{\left( t \right)},\boldsymbol{\theta }^{\left[ k \right]} \right] &=\sum_{m=1}^M{\left| a_m \right|^2w_{m}^{\left( t \right)}},
		\\
		\mathbb{D}\left[ s^{\left( t \right)}|\boldsymbol{y}^{\left( t \right)},\boldsymbol{\theta }^{\left[ k \right]} \right] &=\mathbb{E}\left[ \left| s^{\left( t \right)} \right|^2|\boldsymbol{y}^{\left( t \right)},\boldsymbol{\theta }^{\left[ k \right]} \right] 
		\\
		&\ \ \ \ \ \ \ \ \ \ \ \ \ \ -\left| \mathbb{E}\left[ s^{\left( t \right)}|\boldsymbol{y}^{\left( t \right)},\boldsymbol{\theta }^{\left[ k \right]} \right] \right|^2,
	\end{aligned}
\end{equation}
where $\hat{s}^{\left( t \right)}=\mathbb{E}\left[ s^{\left( t \right)}|\boldsymbol{y}^{\left( t \right)},\boldsymbol{\theta }^{\left[ k \right]} \right]$, $u^{\left( t \right)}=\mathbb{E}\left[ \left| s^{\left( t \right)} \right|^2|\boldsymbol{y}^{\left( t \right)},\boldsymbol{\theta }^{\left[ k \right]} \right]$ and $v^{\left( t \right)}=\mathbb{D}\left[ s^{\left( t \right)}|\boldsymbol{y}^{\left( t \right)},\boldsymbol{\theta }^{\left[ k \right]} \right] $. By aggregating these quantities over all $T$ symbols, we form the following updated vectors
\begin{equation}\label{m2_equation}
	\begin{aligned}
\boldsymbol{s}^{\left[ k \right]}&=\left[ \hat{s}^{\left( 1 \right)},\hat{s}^{\left( 2 \right)},\cdots ,\hat{s}^{\left( T \right)} \right] ,
\\
\boldsymbol{u}^{\left[ k \right]}&=\left[ {u}^{\left( 1 \right)},{u}^{\left( 2 \right)},\cdots ,{u}^{\left( T \right)} \right] ,
\\
\boldsymbol{v}^{\left[ k \right]}&=\left[ {v}^{\left( 1 \right)},{v}^{\left( 2 \right)},\cdots ,{v}^{\left( T \right)} \right] .
	\end{aligned}
\end{equation}

Constructing the Q-function, we have

\begin{equation}
	\begin{aligned}
		&Q\left( \boldsymbol{\theta |\theta }^{\left[ k \right]} \right) =\mathbb{E}_{\boldsymbol{S|Y,\theta }^{\left[ k \right]}}\left[ \ln \mathbb{P}\left[ \boldsymbol{Y,S|\theta } \right] \right] 
		\\
		&=-\sum_{t=1}^T{\mathbb{E}_{\boldsymbol{S|Y,\theta }^{\left[ k \right]}}\left[ \left( \boldsymbol{y}^{\left( t \right)}-\boldsymbol{h}s^{\left( t \right)} \right) ^H\boldsymbol{\Sigma }  ^{-1}\left( \boldsymbol{y}^{\left( t \right)}-\boldsymbol{h}s^{\left( t \right)} \right) \right]}
		\\
		&\ \ \ \ - T\ln \left| \boldsymbol{\Sigma }  \right|+C
		\\
		&=
		-\ln \left| \boldsymbol{\Sigma } \right|-\boldsymbol{h}_{s}^{H}\boldsymbol{Ah}_s\sum_{t=1}^T{u^{\left( t \right)}}+C
		\\
		&\ \ \ \ -\sum_{t=1}^T{\left[ \left( \boldsymbol{y}_{s}^{\left( t \right)} \right) ^H\boldsymbol{Ay}_{s}^{\left( t \right)}+\left( \boldsymbol{y}_{n}^{\left( t \right)} \right) ^H\boldsymbol{Cy}_{n}^{\left( t \right)} \right.}
		\\
		&\ \ \ \ \left. +\left( \boldsymbol{y}_{s}^{\left( t \right)} \right) ^H\boldsymbol{By}_{n}^{\left( t \right)}+\left( \boldsymbol{y}_{n}^{\left( t \right)} \right) ^H\boldsymbol{B}^H\boldsymbol{y}_{s}^{\left( t \right)} \right] 
		\\
		&\ \ \ \ -\sum_{t=1}^T{\left[ \left( \hat{s}^{\left( t \right)} \right) ^*\boldsymbol{h}_{s}^{H}\boldsymbol{Ay}_{s}^{\left( t \right)} \right. +\hat{s}^{\left( t \right)}\left( \boldsymbol{y}_{s}^{\left( t \right)} \right) ^H\boldsymbol{A}^H\boldsymbol{h}_s}
		\\
		&\ \ \ \ +\left. \left( \hat{s}^{\left( t \right)} \right) ^*\boldsymbol{h}_{s}^{H}\boldsymbol{By}_{n}^{\left( t \right)}+\hat{s}^{\left( t \right)}\left( \boldsymbol{y}_{n}^{\left( t \right)} \right) ^H\boldsymbol{B}^H\boldsymbol{h}_s \right] 		
	\end{aligned}
\end{equation}
where $C$ represents terms independent of $\boldsymbol{\theta  }$. Expanding the quadratic form and substituting the conditional moments yields a more explicit expression suitable for optimization in the maximization step.

\subsubsection{Maximization Step}

In the maximization step, we update the parameter estimates by maximizing the Q-function with respect to $\boldsymbol{\theta }=\left\{ \boldsymbol{h}_s,\boldsymbol{\Sigma} \right\} $.

The channel gain vector $\boldsymbol{h}_s$ is updated by setting $\frac{\partial Q\left( \boldsymbol{\theta |\theta }^{\left[ k \right]} \right)}{\partial \boldsymbol{h}_s^*}=0$, we have 
\begin{equation}
	\begin{aligned}
		\boldsymbol{A}^{\left[k \right] }\sum_{t=1}^T{\boldsymbol{y}_{s}^{\left( t \right)}\hat{s}^{\left( t \right) *}}&-\boldsymbol{A}^{\left[k \right] }\bm{h}_s\sum_{t=1}^T{\mathbb{E}\left[ \left| s^{\left( t \right)} \right|^2|\boldsymbol{y}^{\left( t \right)},\boldsymbol{\theta }^{\left[ k \right]} \right]}
		\\
		&\ \ \ \ \ \ \ \ \ \ \ \ \ \ \ \ \ \ \ +\boldsymbol{B}^{\left[k \right] }\sum_{t=1}^T{\boldsymbol{y}_{n}^{\left( t \right)}\hat{s}^{\left( t \right) *}}=0.
	\end{aligned}
\end{equation}

Solving for $\boldsymbol{h}_s$ gives the closed-form update
\begin{equation}\label{hs_equation}
	\begin{aligned}
		\boldsymbol{h}_{s}^{\left[ k+1 \right]}&=\frac{\sum_{t=1}^T{\boldsymbol{y}_{s}^{\left( t \right)}\hat{s}^{\left( t \right) *}}+\left( \boldsymbol{A}^{\left[ k \right]} \right) ^{-1}\boldsymbol{B}^{\left[ k \right]}\sum_{t=1}^T{\boldsymbol{y}_{n}^{\left( t \right)}\hat{s}^{\left( t \right) *}}}{\sum_{t=1}^T{\mathbb{E}\left[ \left| s^{\left( t \right)} \right|^2|\boldsymbol{y}^{\left( t \right)},\boldsymbol{\theta }^{\left[ k \right]} \right]}}
		\\
		&=\frac{\sum_{t=1}^T{\boldsymbol{y}_{s}^{\left( t \right)}\hat{s}^{\left( t \right) *}}-\boldsymbol{\Sigma }_{sn}^{\left[ k \right]}\left( \boldsymbol{\Sigma }_{nn}^{\left[ k \right]} \right) ^{-1}\sum_{t=1}^T{\boldsymbol{y}_{n}^{\left( t \right)}\hat{s}^{\left( t \right) *}}}{\sum_{t=1}^T{u^{\left( t \right)} }},
	\end{aligned}
\end{equation}
where $\boldsymbol{h}_{s}^{\left[ k+1 \right]}=\frac{\sum_{t=1}^T{\boldsymbol{y}_{s}^{\left( t \right)}\hat{s}^{\left( t \right) *}}}{\sum_{t=1}^T{u^{\left( t \right)} }}$ for $N_n=0$.

The channel covariance matrix $\boldsymbol{\Sigma }$ is updated by setting $\frac{\partial Q\left( \boldsymbol{\theta |\theta }^{\left[ k \right]} \right)}{\partial \bm{\Sigma}}=0$, we have 

\begin{equation}\label{S_equation}
	\begin{aligned}
		\boldsymbol{\Sigma }^{\left[ k+1 \right]}&=\frac{1}{T}\sum_{t=1}^T{\mathbb{E}\left[ \left( \boldsymbol{y}^{\left( t \right)}-\boldsymbol{h}^{\left[ k+1 \right]}s^{\left( t \right)} \right) \right.}
		\\
		&\ \ \ \ \ \ \ \ \ \ \ \ \times \left. \left( \boldsymbol{y}^{\left( t \right)}-\boldsymbol{h}^{\left[ k+1 \right]}s^{\left( t \right)} \right) ^H|\boldsymbol{y}^{\left( t \right)},\boldsymbol{\theta }^{\left[ k \right]} \right] 
		\\
		&=\frac{1}{T}\sum_{t=1}^T{\left[ v^{\left( t \right)}\boldsymbol{h}^{\left[ k+1 \right]}\left( \boldsymbol{h}^{\left[ k+1 \right]} \right) ^H \right.}
		\\
		&\ \ \ \ \ \ \  \left. +\left( \boldsymbol{y}^{\left( t \right)}-\boldsymbol{h}^{\left[ k+1 \right]}\hat{s}^{\left( t \right)} \right) \left( \boldsymbol{y}^{\left( t \right)}-\boldsymbol{h}^{\left[ k+1 \right]}\hat{s}^{\left( t \right)} \right) ^H \right] ,
	\end{aligned}
\end{equation}
where we obtain the block recursive formula for the block covariance matrix by substituting $ \boldsymbol{y}^{(t)} = \begin{bmatrix} \boldsymbol{y}_s^{(t)} \\ \boldsymbol{y}_n^{(t)} \end{bmatrix}$ and $\boldsymbol{h}=\left[ \begin{array}{c}
	\boldsymbol{h}_s\\
	\boldsymbol{h}_n\\
\end{array} \right]$ into Eq. (\ref{S_equation}), given by

\begin{equation}\label{Ss_equation}
	\begin{aligned}
\boldsymbol{\Sigma }_{ss}^{\left[ k+1 \right]}&=\frac{1}{T}\sum_{t=1}^T{\left[ \left( \boldsymbol{y}_{s}^{\left( t \right)}-\boldsymbol{h}_{s}^{\left[ k+1 \right]}\hat{s}^{\left( t \right)} \right) \left( \boldsymbol{y}_{s}^{\left( t \right)}-\boldsymbol{h}_{s}^{\left[ k+1 \right]}\hat{s}^{\left( t \right)} \right) ^H \right.}
\\
&\ \ \ \ \ \ \ \ \ \ \ \ \ \left. +\boldsymbol{h}_{s}^{\left[ k+1 \right]}\left( \boldsymbol{h}_{s}^{\left[ k+1 \right]} \right) ^H\mathbb{D}\left[ s^{\left( t \right)}|\boldsymbol{y}^{\left( t \right)},\boldsymbol{\theta }^{\left[ k \right]} \right] \right] ,
\\
\boldsymbol{\Sigma }_{sn}^{\left[ k+1 \right]}&=\frac{1}{T}\sum_{t=1}^T{\left[ \left( \boldsymbol{y}_{s}^{\left( t \right)}-\boldsymbol{h}_{s}^{\left[ k+1 \right]}\hat{s}^{\left( t \right)} \right) \left( \boldsymbol{y}_{n}^{\left( t \right)} \right) ^H \right]},
	\end{aligned}
\end{equation}
where $\boldsymbol{\Sigma }_{ss}^{\left[ k+1 \right]}$ and $\boldsymbol{\Sigma }_{sn}^{\left[ k+1 \right]}$ require iterative updates, while $\boldsymbol{\Sigma }_{nn}$ is estimated once from the noise-only observations, given by

\begin{equation}\label{Snn_equation}
	\begin{aligned}
\boldsymbol{\Sigma }_{nn}&=\frac{1}{T}\sum_{t=1}^T{\left[ \boldsymbol{y}_{n}^{\left( t \right)}\left( \boldsymbol{y}_{n}^{\left( t \right)} \right) ^H \right]},
	\end{aligned}
\end{equation}
where $\boldsymbol{\Sigma }_{nn}=\boldsymbol{0}$ and $\boldsymbol{\Sigma }_{sn}^{\left[ k+1 \right]}=\boldsymbol{0}$ for $N_n=0$. And we enforce hermitian symmetry to ensure convergence with $\boldsymbol{\Sigma }_{ss}^{\left[ k+1 \right]}=\frac{\boldsymbol{\Sigma }_{ss}^{\left[ k+1 \right]}+\left( \boldsymbol{\Sigma }_{ss}^{\left[ k+1 \right]} \right) ^H}{2}$ and $\boldsymbol{\Sigma }_{sn}^{\left[ k+1 \right]}=\frac{\boldsymbol{\Sigma }_{sn}^{\left[ k+1 \right]}+\left( \boldsymbol{\Sigma }_{sn}^{\left[ k+1 \right]} \right) ^H}{2}$ after each iteration.

\subsubsection{Calibration Step}

Due to the inherent scale ambiguity in the maximum likelihood EM algorithm for estimating $\boldsymbol{h}$ and $s$, the model $\boldsymbol{y} = \boldsymbol{h} s + \boldsymbol{n}$ is invariant under the transformation $(\boldsymbol{h}, s) \rightarrow (\alpha \boldsymbol{h}, s / \alpha)$ for any nonzero scalar $\alpha$. This means that different combinations of channel gains and transmitted symbols can yield identical received observations. Consequently, the EM algorithm may converge to estimates $\hat{\boldsymbol{h}}$ and $\hat{s}$ that differ from the true values by an unknown scale factor $\alpha$, although their product $\hat{\boldsymbol{h}} \hat{s}$ remains correct.

 To resolve this ambiguity, we terminate the EM algorithm when the changes in $\hat{\boldsymbol{h}} \hat{s}$ and the covariance matrix $\bm{\Sigma}$ fall below thresholds of $\epsilon_{\bm{hs}} $ and $\epsilon_{\bm{\Sigma}} $, respectively. We then calibrate the estimated symbols using the average power of the QAM constellation. The scaling factor $\alpha$ and the calibrated symbols $\boldsymbol{\hat{s}}$ can be expressed as
\begin{equation}\label{Cal_equation}
	\begin{aligned}
\boldsymbol{\hat{s}}=\sqrt{\frac{2T\left( M-1 \right)}{3\sum_{t=1}^T{\left| \hat{s}^{\left( t \right)} \right|^2}}}\boldsymbol{s}^{\left[ k \right]},
	\end{aligned}
\end{equation}
where $\boldsymbol{s}^{[k]}$ and $\hat{s}^{\left(t \right) }$ denote the uncalibrated symbol estimates obtained after convergence of the EM algorithm following the $k$-th iteration, and $\boldsymbol{\hat{s}}$ represents the calibrated symbols. The factor $\sqrt{2T(M-1) / (3 \sum_{t=1}^T |\hat{s}^{(t)}|^2)}$ incorporates both the QAM constellation power $P_s=\frac{2}{3}(M-1)$ and the scaling needed to match the average power of the transmitted symbols.

Using the calibrated symbols ${\hat{\boldsymbol{s}}}$, we recompute the calibrated channel gains ${\boldsymbol{\hat{h}}}_s$ and covariance matrix ${\boldsymbol{\hat{\Sigma} }}_{ss}$, ${\boldsymbol{\hat{\Sigma} }}_{sn}$, ${\boldsymbol{\hat{\Sigma} }_{nn}}$. Based on these calibrated parameters, the final detected symbol $\boldsymbol{\bar{s}}=\left\{ \bar{s}^{\left( 1 \right)},\cdots ,\bar{s}^{\left( T \right)} \right\} $ and $\bar{s}^{(t)}=a_{m^*}$ is obtained either by selecting the symbol with the maximum posterior probability 
\begin{equation}\label{detect1_equation}
	\begin{aligned}
		m^* = \arg\max_{m} w_{m}^{(t)},
	\end{aligned}
\end{equation}
or by minimizing the distance to the constellation point
\begin{equation}\label{detect2_equation}
	\begin{aligned}
	m^* = \arg\min_{m} |\hat{s}^{\left( t\right) } - a_m|.
	\end{aligned}
\end{equation}

\begin{algorithm}
	\caption{}
	\label{EM2}
	\begin{algorithmic}[1]
		\Require $\boldsymbol{Y}=\left[ \boldsymbol{Y}_s;\boldsymbol{Y}_n \right] $, $\mathcal{A}$, $\epsilon_{\bm{hs}} > 0$, $\epsilon_{\bm{\Sigma}} > 0$;
		\State Initialize: $\boldsymbol{h}_s^{\left[0 \right] }$, $\boldsymbol{\Sigma}_{ss}^{\left[0 \right] }$, $\boldsymbol{\Sigma}_{sn}^{\left[0 \right] }$;
		\State Calculate $\boldsymbol{\Sigma}_{nn}$ based on $\boldsymbol{Y}_n$ according to Eq. (\ref{Snn_equation});
		
		\Repeat
		\State \textbf{Expectation Step}:
		\State Calculate $\bm{A}^{\left[ k \right]}$, $\bm{B}^{\left[ k \right]}$, and $\bm{C}^{\left[ k \right]}$ based on $\boldsymbol{\Sigma}_{ss}^{\left[ k \right]}$, $\boldsymbol{\Sigma}_{sn}^{\left[ k \right]}$ and $\boldsymbol{\Sigma }_{nn}$ according to Eq. (\ref{ABC_equation});

		\State Calculate $d_{m}^{\left( t \right)}$ based on $\boldsymbol{Y}$, $\mathcal{A}$, $\boldsymbol{h}_s^{\left[ k \right]}$, $\bm{A}^{\left[ k \right]}$, $\bm{B}^{\left[ k \right]}$, and $\bm{C}^{\left[ k \right]}$ according to Eq. (\ref{d_equation}) for all $m = 1,\cdots,M$ and $t=1,\cdots,T$;
		\State Calculate $w_{m}^{\left( t \right)}$ based on $d_{m}^{\left( t \right)}$ according to Eq. (\ref{w_equation}) for all $m = 1,\cdots,M$ and $t=1,\cdots,T$;

		\State Calculate $\boldsymbol{s}^{\left[ k \right]}$, $\boldsymbol{u}^{\left[ k \right]}$ and $\boldsymbol{v}^{\left[ k \right]}$ based on $\mathcal{A}$ and $w_{m}^{\left( t \right)}$ according to Eq. (\ref{m_equation}) and Eq. (\ref{m2_equation}) for all $t=1,\cdots,T$;
		
		\State \textbf{Maximization Step}:
		\State Calculate $\boldsymbol{h}_s^{\left[ k+1 \right]}$ based on $\boldsymbol{Y}$, $\boldsymbol{s}^{\left[ k \right]}$, $\boldsymbol{u}^{\left[ k \right]}$, $\boldsymbol{\Sigma}_{sn}^{\left[ k \right]}$ and $\boldsymbol{\Sigma }_{nn}$ according to Eq. (\ref{hs_equation});
		\State Calculate $\boldsymbol{\Sigma}_{ss}^{\left[ k+1 \right]}$ and $\boldsymbol{\Sigma}_{sn}^{\left[ k+1 \right]}$ based on $\boldsymbol{Y}$, $\boldsymbol{h}_s^{\left[ k+1 \right]}$, $\boldsymbol{s}^{\left[ k \right]}$, $\boldsymbol{v}^{\left[ k \right]}$ according to Eq. (\ref{S_equation});
		
		\Until{$\lVert \boldsymbol{h}^{\left[ k+1 \right]}\boldsymbol{s}^{\left[ k+1 \right]}-\boldsymbol{h}^{\left[ k \right]}\boldsymbol{s}^{\left[ k \right]} \rVert _F<\epsilon _{\bm{hs}}$ and $\lVert \boldsymbol{\Sigma }^{\left[ k+1 \right]}-\boldsymbol{\Sigma }^{\left[ k \right]} \rVert _F<\epsilon _{\bm{\Sigma}}$};
		
		\State \textbf{Calibration Step}:
		\State Calibrate ${\boldsymbol{\hat{s}}}$ based on $ \boldsymbol{s}^{\left[ k \right]}$ according to Eq. (\ref{Cal_equation});
		\State Calibrate $\hat{\boldsymbol{h}}_s$ based on $\boldsymbol{Y}$, ${\boldsymbol{\hat{s}}}$, $\boldsymbol{u}^{\left[ k \right]}$, $\boldsymbol{\Sigma }_{sn}^{\left[ k \right]}$ and $\boldsymbol{\Sigma }_{nn}$ according to Eq. (\ref{hs_equation});
		\State Calibrate ${\boldsymbol{\hat{\Sigma} }}_{ss}=\boldsymbol{\Sigma }_{ss}^{\left[ k \right]}$, ${\boldsymbol{\hat{\Sigma} }}_{sn}=\boldsymbol{\Sigma }_{sn}^{\left[ k \right]}$, and ${\boldsymbol{\hat{\Sigma} }}_{nn}=\boldsymbol{\Sigma }_{nn}$;
		\State Calculate $\bm{\bar{s}}$ based on $\hat{\boldsymbol{s}}$ according to Eq. (\ref{detect1_equation}) and Eq. (\ref{detect2_equation});
		\State Output: $\boldsymbol{\hat{\bm{s}}}$, $\boldsymbol{\bar{s}}$, $\hat{\boldsymbol{h}}_s$, $\hat{\boldsymbol{\Sigma }}_{ss}$, $\hat{\boldsymbol{\Sigma }}_{sn}$ and $\hat{\boldsymbol{\Sigma }_{nn}}$.
	\end{algorithmic}
\end{algorithm}

\subsection{Complexity}\label{T42}

For the $ k $-th iteration of the proposed EM algorithm, the computational complexity of the expectation step is primarily dominated by the computation of the posterior probability for each constellation point, which scales as $\mathcal{O}(M N N)$. For $ T $ symbols, the total complexity of the expectation step is $\mathcal{O}(T M N^2) $. The complexity of the maximization step is dominated by the update of the covariance matrix, which scales as $ \mathcal{O}(T N^2) $. Therefore, for a total of $ T $ symbols over $ K $ iterations, the overall complexity of the EM algorithm is $ \mathcal{O}(K T M N^2) $.

\section{RESULTS}\label{T5}

\begin{figure}
	\centering
	\includegraphics[width=0.45\textwidth]{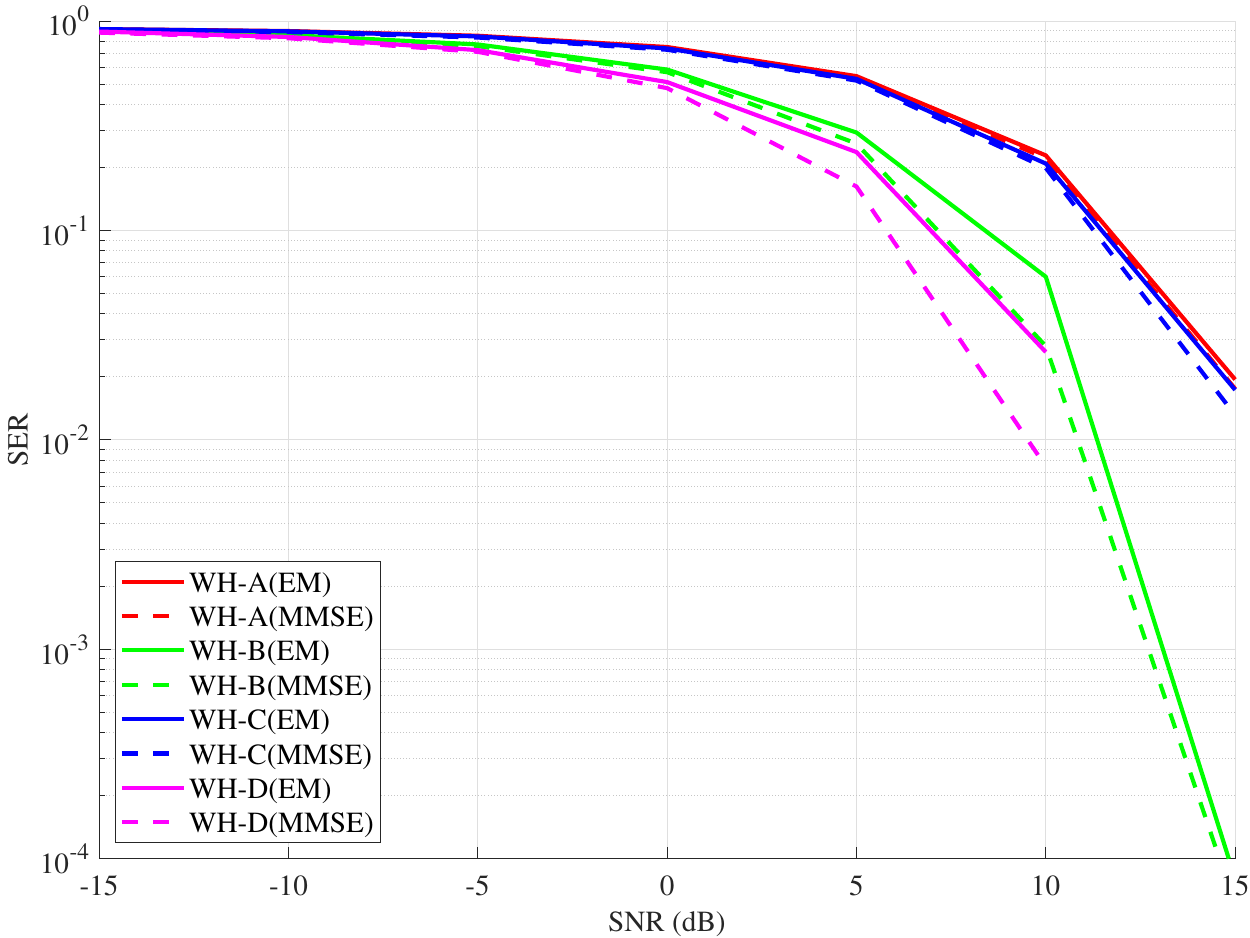}
	\caption{SER performance under $M=16$ and $T=10^5$.}
	\label{Mat_BER_M16_SNR_15_15_T100}
\end{figure}

\begin{figure}
	\centering
	\includegraphics[width=0.45\textwidth]{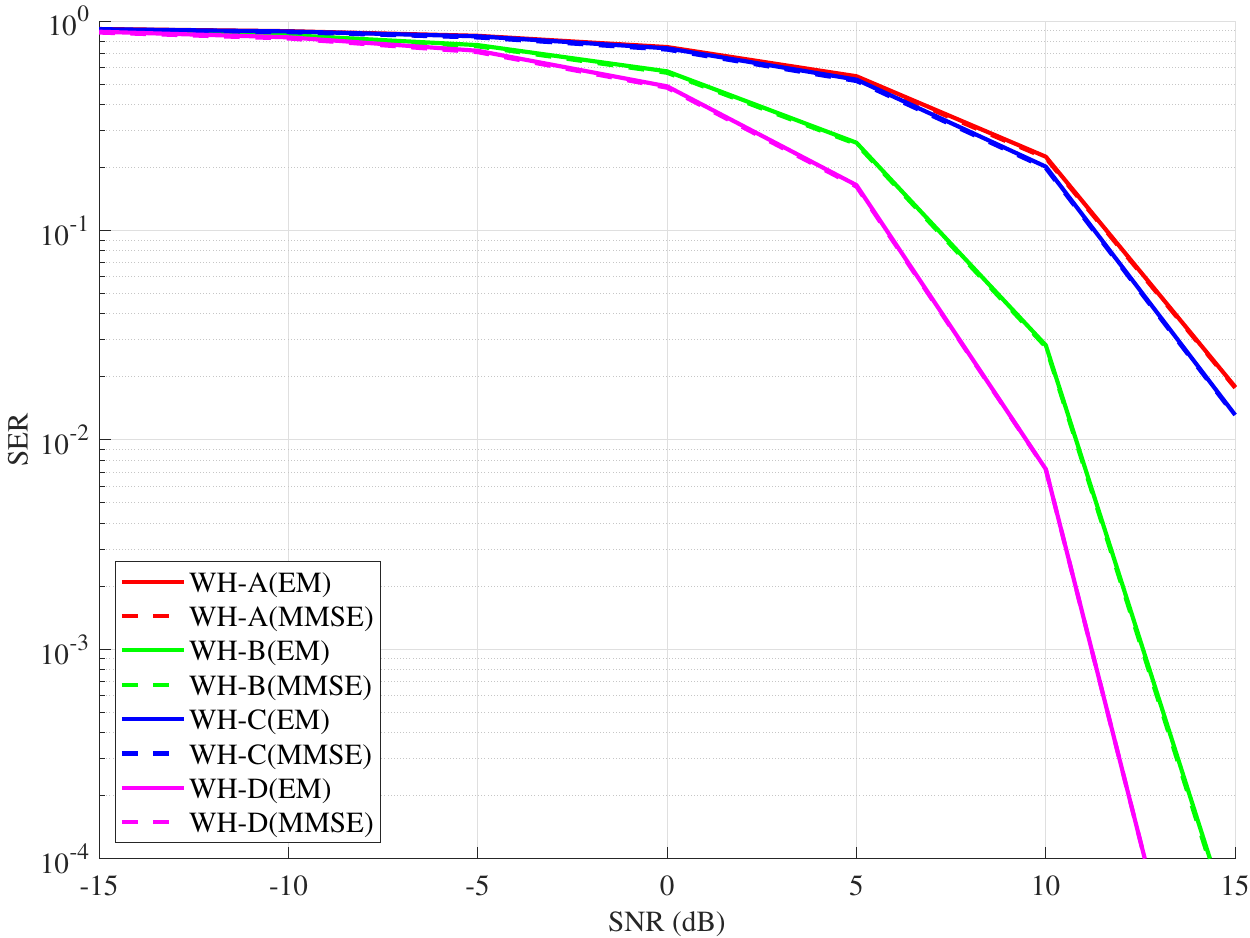}
	\caption{SER performance under $M=16$ and $T=10^2$.}
	\label{Mat_BER_M16_SNR_15_15_T100000}
\end{figure}

\begin{figure}
	\centering
	\includegraphics[width=0.45\textwidth]{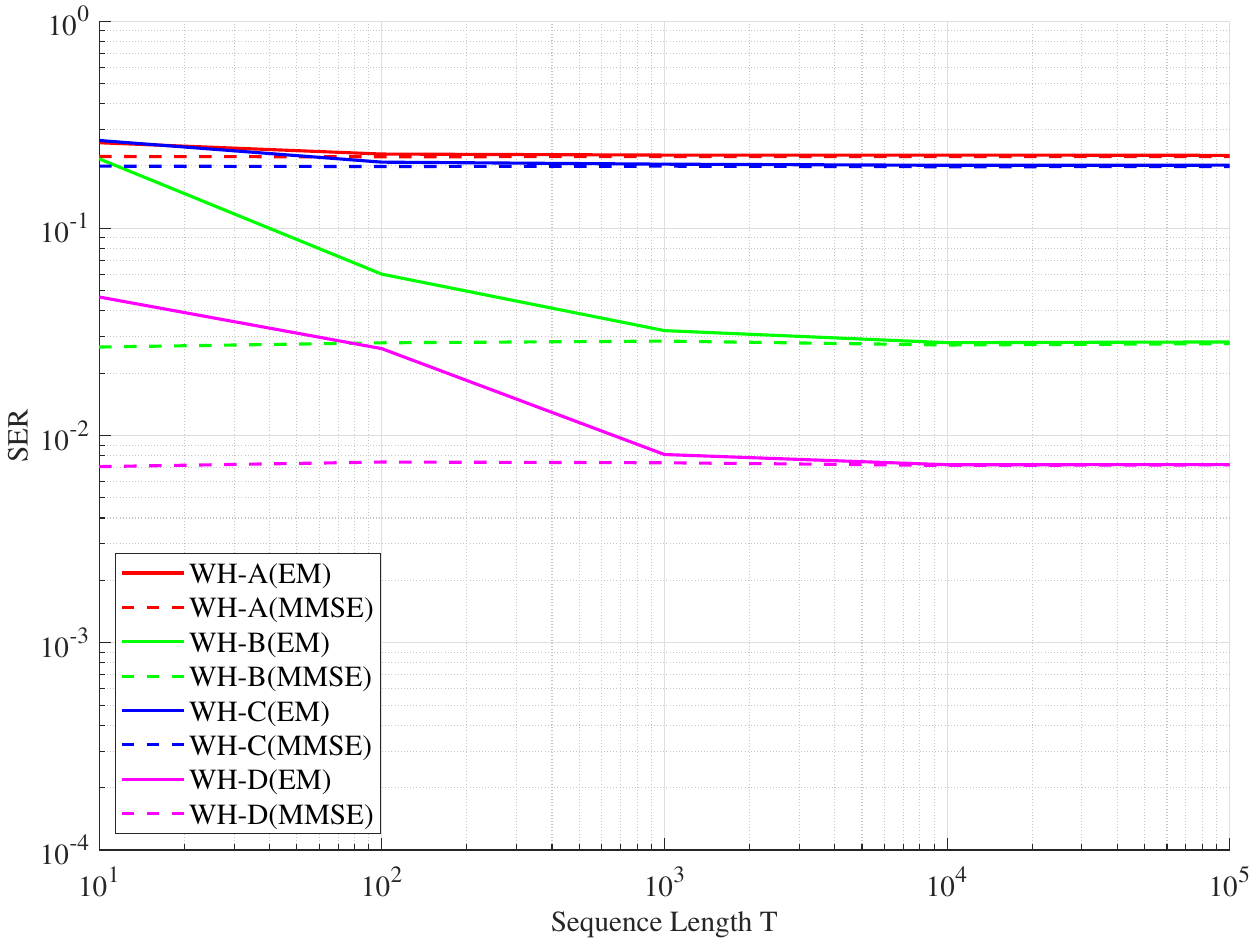}
	\caption{SER performance under $M=16$ and $SNR=10$.}
	\label{Mat_SER_M16_SNR_10_T10_100000}
\end{figure}

Figures \ref{Mat_BER_M16_SNR_15_15_T100} and \ref{Mat_BER_M16_SNR_15_15_T100000} compare the SER performance under varying sequence lengths. The MMSE algorithm assumes known overall system gain and noise covariance matrix (Section \ref{T3}), whereas the EM algorithm operates under unknown gain and noise covariance matrix (Section \ref{T4}). The MMSE algorithm performs symbol-by-symbol decoding, while the EM algorithm performs joint estimation over a sequence of $T$ symbols. Figure \ref{Mat_SER_M16_SNR_10_T10_100000} demonstrates the relationship between sequence length $T$ and SER performance. It is evident that as $T$ increases, the SER performance of the EM algorithm asymptotically approaches that of the MMSE algorithm with perfect knowledge of gain and noise covariance.

\section{Extensions and Future Directions}\label{T6}

From the perspective of extensibility, the WH architecture can be readily adapted to other complex Rydberg sensing frameworks. The WH architecture is motivated by the principle that fusing information from multiple channels consistently outperforms single-channel approaches. Based on this insight, variants of the WH framework can be broadly applied to a range of complex Rydberg sensing architectures to achieve improved performance. For example, although fluorescence-based Rydberg detection offers low background noise, it still relies on probe and coupling lasers that carry useful information; thus, a WH extension for such architectures would incorporate fluorescence signals alongside laser outputs. Similarly, in multi-laser excitation schemes, each laser beam generally contains partial information about the measured field, so fusing more channels yields better performance.

From a commercialization perspective, conventional two-laser Rydberg architectures typically rely on expensive, high-precision lasers, which limits their commercial viability. In contrast, three-laser excitation schemes offer greater cost-effectiveness due to the use of more affordable lasers, making them more suitable for commercial deployment. By integrating the WH architecture with two- or three-laser Rydberg sensing, it becomes possible to deploy low-cost, lower-performance lasers, such as LED-based sources, in future Rydberg sensing systems, further enhancing their potential for commercial adoption.

From the perspective of compatibility with other signal processing algorithms, the WH architecture is not limited to the specific implementation presented in this work. While our current framework is based on matched filtering and symbol detection for each channel followed by maximum likelihood combining, other fusion strategies can also be applied. For instance, adaptive filtering techniques such as Wiener filtering, least mean squares (LMS), or recursive least squares (RLS) algorithms can be employed to jointly process multi-channel inputs, potentially achieving further performance gains. This flexibility underscores the generality of the WH concept and its adaptability to various signal processing paradigms in Rydberg sensing.

From the perspective of sensitivity improvement in Rydberg architectures, the WH framework and its variants effectively mitigate hardware-induced noise, particularly from lasers, thereby offering, in principle, better signal-to-noise ratios than single-probe-channel schemes. However, since the weighting process is performed on electrical signals after ADC sampling rather than on optical signals, the additional noise introduced by ADCs must be considered. Because ADC noise is uncorrelated across channels, it cannot be canceled by the WH architecture. Nevertheless, for Rydberg receivers targeting communication applications, ADC sampling is an indispensable part of the signal chain. Therefore, despite this limitation, the WH architecture remains well suited for communication receivers, as the overall performance gain from multi-channel weighting outweighs the uncorrelated ADC noise floor.

From the perspective of array deployment, the design principles of the WH architecture can also be applied to Rydberg receiver arrays. Currently, such arrays often use a high-power laser split into multiple beams, each serving as a small antenna element. Because these antennas share the same laser source, the laser-induced noise across different elements is typically correlated. This correlation enables a shared noise reference channel to be used for canceling common-mode noise across multiple antennas, following the same logic as the WH framework. By extending the WH architecture to array configurations, it becomes possible to jointly process signals from multiple Rydberg antennas while using only a single or a few noise reference channels, thereby improving detection performance without proportionally increasing hardware complexity.

\section{Conclusion} \label{T7}

This paper proposed a weight hybrid architecture for Rydberg atomic receivers to mitigate hardware-induced correlated noise, particularly from lasers. By jointly processing dual signal and noise reference channels, the WH architecture moves beyond conventional single-probe readout. For known channel parameters, we derived optimal weighting coefficients and quantified SNR improvement. For unknown parameters, we developed an MLE-based algorithm with an improved expectation-maximization framework for robust signal extraction. Experimental results validated the superiority of the proposed approach. The WH architecture is general and readily extendable to other Rydberg sensing systems, offering a practical solution for high-sensitivity atomic radio reception under correlated noise.

%\appendices
%
%
%\section{MLE of Intensity Superheterodyne Detection}\label{A12}
%
%
%
%
%\section{CRLB and Iteration Calculation of Splitting Detection}\label{A3}
%
%\subsection{Iteration Calculation of Univariate Peak Shift Estimation}\label{A31}

\ifCLASSOPTIONcaptionsoff
\newpage
\fi

\normalem
\bibliographystyle{IEEEtran}
\bibliography{myref}

@article{kumar2023quantum,
  title={Quantum-enabled millimetre wave to optical transduction using neutral atoms},
  author={Kumar, Aishwarya and Suleymanzade, Aziza and Stone, Mark and Taneja, Lavanya and Anferov, Alexander and Schuster, David I and Simon, Jonathan},
  journal={Nature},
  volume={615},
  number={7953},
  pages={614--619},
  year={2023},
  publisher={Nature Publishing Group UK London}
}

@article{borowka2024continuous,
  title={Continuous wideband microwave-to-optical converter based on room-temperature Rydberg atoms},
  author={Bor{\'o}wka, Sebastian and Pylypenko, Uliana and Mazelanik, Mateusz and Parniak, Micha{\l}},
  journal={Nature Photonics},
  volume={18},
  number={1},
  pages={32--38},
  year={2024},
  publisher={Nature Publishing Group UK London}
}

@article{manchaiah2026probing,
  title={Probing bandwidth and sensitivity in Rydberg atom sensing via optical homodyne and rf heterodyne detection},
  author={Manchaiah, Dixith and Oliver, Stone and Berweger, Samuel and Holloway, Christopher L and Prajapati, Nikunjkumar},
  journal={Physical Review A},
  volume={113},
  number={1},
  pages={013729},
  year={2026},
  publisher={APS}
}

@article{jiang2025quantum,
  title={Quantum weak measurement amplifies dispersion signal of Rydberg atomic system},
  author={Jiang, Yinghang and Wu, Jiguo and Shi, Meng and Zheng, Hanqing and Guo, Fei and Xiao, Zhiguang and Zhang, Zhiyou},
  journal={Communications Physics},
  volume={8},
  number={1},
  pages={144},
  year={2025},
  publisher={Nature Publishing Group UK London}
}

@article{liang2026exceptional,
  title={Exceptional Point-Enhanced Rydberg Atomic Electrometers},
  author={Liang, Chao and Yang, Ce and Huang, Wei and You, Li},
  journal={Physical Review Letters},
  volume={136},
  number={5},
  pages={053203},
  year={2026},
  publisher={APS}
}

@article{knarr2023spatiotemporal,
  title={Spatiotemporal multiplexed Rydberg receiver},
  author={Knarr, Samuel H and Bucklew, Victor G and Langston, Jerrod and Cox, Kevin C and Hill, Joshua C and Meyer, David H and Drakes, James A},
  journal={IEEE Transactions on Quantum Engineering},
  volume={4},
  pages={1--8},
  year={2023},
  publisher={IEEE}
}

@article{qimeng2025instantaneous,
  title={Instantaneous Bandwidth Expansion of a Gradient Magnetic Field Enhanced Rydberg Atomic Receiver},
  author={Qimeng, Wang and An, Qiang and Fu, Yunqi},
  journal={IEEE Sensors Journal},
  year={2025},
  publisher={IEEE}
}

@article{prajapati2023sensitivity,
  title={Sensitivity comparison of two-photon vs three-photon Rydberg electrometry},
  author={Prajapati, Nikunjkumar and Bhusal, Narayan and Rotunno, Andrew P and Berweger, Samuel and Simons, Matthew T and Artusio-Glimpse, Alexandra B and Ju Wang, Ying and Bottomley, Eric and Fan, Haoquan and Holloway, Christopher L},
  journal={Journal of Applied Physics},
  volume={134},
  number={2},
  year={2023},
  publisher={AIP Publishing}
}

@article{prajapati2024high,
  title={High angular momentum coupling for enhanced Rydberg-atom sensing in the very-high frequency band},
  author={Prajapati, Nikunjkumar and Kunzler, Jakob W and Artusio-Glimpse, Alexandra B and Rotunno, Andrew P and Berweger, Samuel and Simons, Matthew T and Holloway, Christopher L and Gardner, Chad M and Mcbeth, Michael S and Younts, Robert A},
  journal={Journal of Applied Physics},
  volume={135},
  number={7},
  year={2024},
  publisher={AIP Publishing}
}

@article{chen2022terahertz,
  title={Terahertz electrometry via infrared spectroscopy of atomic vapor},
  author={Chen, Shuying and Reed, Dominic J and MacKellar, Andrew R and Downes, Lucy A and Almuhawish, Nourah FA and Jamieson, Matthew J and Adams, Charles S and Weatherill, Kevin J},
  journal={Optica},
  volume={9},
  number={5},
  pages={485--491},
  year={2022},
  publisher={Optica Publishing Group}
}

@article{brown2023very,
  title={Very-high-and ultrahigh-frequency electric-field detection using high angular momentum Rydberg states},
  author={Brown, Roger C and Kayim, Baran and Viray, Michael A and Perry, Abigail R and Sawyer, Brian C and Wyllie, Robert},
  journal={Physical Review A},
  volume={107},
  number={5},
  pages={052605},
  year={2023},
  publisher={APS}
}

@article{bohaichuk2023three,
  title={Three-photon Rydberg-atom-based radio-frequency sensing scheme with narrow linewidth},
  author={Bohaichuk, Stephanie M and Ripka, Fabian and Venu, Vijin and Christaller, Florian and Liu, Chang and Schmidt, Matthias and K{\"u}bler, Harald and Shaffer, James P},
  journal={Physical Review Applied},
  volume={20},
  number={6},
  pages={L061004},
  year={2023},
  publisher={APS}
}

@article{johnson2012three,
  title={A three-step laser stabilization scheme for excitation to Rydberg levels in 85Rb},
  author={Johnson, LAM and Majeed, HO and Varcoe, BTH},
  journal={Applied Physics B},
  volume={106},
  number={2},
  pages={257--260},
  year={2012},
  publisher={Springer}
}

@article{fahey2011excitation,
  title={Excitation of Rydberg states in rubidium with near infrared diode lasers},
  author={Fahey, Donald P and Noel, Michael W},
  journal={Optics express},
  volume={19},
  number={18},
  pages={17002--17012},
  year={2011},
  publisher={Optical Society of America}
}

@article{you2022microwave,
  title={Microwave-field sensing via electromagnetically induced absorption of Rb irradiated by three-color infrared lasers},
  author={You, SH and Cai, MH and Zhang, SS and Xu, ZS and Liu, HP},
  journal={Optics Express},
  volume={30},
  number={10},
  pages={16619--16629},
  year={2022},
  publisher={Optica Publishing Group}
}

@article{zhang2022rydberg,
  title={Rydberg microwave-frequency-comb spectrometer},
  author={Zhang, Li-Hua and Liu, Zong-Kai and Liu, Bang and Zhang, Zheng-Yuan and Guo, Guang-Can and Ding, Dong-Sheng and Shi, Bao-Sen},
  journal={Physical Review Applied},
  volume={18},
  number={1},
  pages={014033},
  year={2022},
  publisher={APS}
}

@article{thaicharoen2019electromagnetically,
  title={Electromagnetically induced transparency, absorption, and microwave-field sensing in a Rb vapor cell with a three-color all-infrared laser system},
  author={Thaicharoen, N and Moore, KR and Anderson, DA and Powel, RC and Peterson, E and Raithel, G},
  journal={Physical Review A},
  volume={100},
  number={6},
  pages={063427},
  year={2019},
  publisher={APS}
}

@article{carr2012three,
  title={Three-photon electromagnetically induced transparency using Rydberg states},
  author={Carr, Christopher and Tanasittikosol, Monsit and Sargsyan, Armen and Sarkisyan, David and Adams, Charles S and Weatherill, Kevin J},
  journal={Optics letters},
  volume={37},
  number={18},
  pages={3858--3860},
  year={2012},
  publisher={Optical Society of America}
}

@article{johnson2010absolute,
  title={Absolute frequency measurements of 85Rb n F7/2 Rydberg states using purely optical detection},
  author={Johnson, LAM and Majeed, HO and Sanguinetti, B and Becker, Th and Varcoe, BTH},
  journal={New Journal of Physics},
  volume={12},
  number={6},
  pages={063028},
  year={2010}
}

@article{prajapati2024investigation,
  title={Investigation of fluorescence versus transmission readout for three-photon Rydberg excitation used in electrometry},
  author={Prajapati, Nikunjkumar and Berweger, Samuel and Rotunno, Andrew P and Artusio-Glimpse, Alexandra B and Schlossberger, Noah and Shylla, Dangka and Watterson, William J and Simons, Matthew T and LaMantia, David and Norrgard, Eric B and others},
  journal={AVS Quantum Science},
  volume={6},
  number={3},
  year={2024},
  publisher={AIP Publishing}
}

@article{teale2022degenerate,
  title={Degenerate two-photon Rydberg atom voltage reference},
  author={Teale, Carson and Sherman, Jeffrey and Kitching, John},
  journal={AVS quantum science},
  volume={4},
  number={2},
  year={2022},
  publisher={AIP Publishing}
}

@article{holloway2017electric,
  title={Electric field metrology for {SI} traceability: Systematic measurement uncertainties in electromagnetically induced transparency in atomic vapor},
  author={Holloway, Christopher L and Simons, Matt T and Gordon, Joshua A and Dienstfrey, Andrew and Anderson, David A and Raithel, Georg},
  journal={Journal of Applied Physics},
  volume={121},
  number={23},
  pages={233106},
  year={2017},
  publisher={AIP Publishing LLC}
}

@article{robinson2021determining,
  title={Determining the angle-of-arrival of a radio-frequency source with a {Rydberg} atom-based sensor},
  author={Robinson, Amy K and Prajapati, Nikunjkumar and Senic, Damir and Simons, Matthew T and Holloway, Christopher L},
  journal={Applied Physics Letters},
  volume={118},
  number={11},
  pages={114001},
  year={2021},
  publisher={AIP Publishing LLC}
}

@article{jing2020atomic,
  title={Atomic superheterodyne receiver based on microwave-dressed {Rydberg} spectroscopy},
  author={Jing, Mingyong and Hu, Ying and Ma, Jie and Zhang, Hao and Zhang, Linjie and Xiao, Liantuan and Jia, Suotang},
  journal={Nature Physics},
  volume={16},
  number={9},
  pages={911--915},
  year={2020},
  publisher={Nature Publishing Group}
}

@article{holloway2017atom,
  title={Atom-based {RF} electric field metrology: from self-calibrated measurements to subwavelength and near-field imaging},
  author={Holloway, Christopher L and Simons, Matthew T and Gordon, Joshua A and Wilson, Perry F and Cooke, Caitlyn M and Anderson, David A and Raithel, Georg},
  journal={IEEE Transactions on Electromagnetic Compatibility},
  volume={59},
  number={2},
  pages={717--728},
  year={2017},
  publisher={IEEE}
}

@article{downes2020full,
  title={Full-field {Terahertz} imaging at {kiloHertz} frame rates using atomic vapor},
  author={Downes, Lucy A and MacKellar, Andrew R and Whiting, Daniel J and Bourgenot, Cyril and Adams, Charles S and Weatherill, Kevin J},
  journal={Physical Review X},
  volume={10},
  number={1},
  pages={011027},
  year={2020},
  publisher={APS}
}

@article{liu2022deep,
  title={Deep learning enhanced Rydberg multifrequency microwave recognition},
  author={Liu, Zong-Kai and Zhang, Li-Hua and Liu, Bang and Zhang, Zheng-Yuan and Guo, Guang-Can and Ding, Dong-Sheng and Shi, Bao-Sen},
  journal={Nature Communications},
  volume={13},
  number={1},
  pages={1--10},
  year={2022},
  publisher={Nature Publishing Group}
}

@article{sedlacek2013atom,
  title={Atom-based vector microwave electrometry using rubidium {Rydberg} atoms in a vapor cell},
  author={Sedlacek, JA and Schwettmann, A and K{\"u}bler, Harald and Shaffer, JP},
  journal={Physical Review Letters},
  volume={111},
  number={6},
  pages={063001},
  year={2013},
  publisher={APS}
}

@article{holloway2014sub,
  title={Sub-wavelength imaging and field mapping via electromagnetically induced transparency and Autler-Townes splitting in {Rydberg} atoms},
  author={Holloway, Christopher L and Gordon, Joshua A and Schwarzkopf, Andrew and Anderson, David A and Miller, Stephanie A and Thaicharoen, Nithiwadee and Raithel, Georg},
  journal={Applied Physics Letters},
  volume={104},
  number={24},
  pages={244102},
  year={2014},
  publisher={American Institute of Physics}
}

@article{anderson2021self,
  title={A self-calibrated {SI}-traceable {Rydberg} atom-based radio frequency electric field probe and measurement instrument},
  author={Anderson, David Alexander and Sapiro, Rachel Elizabeth and Raithel, Georg},
  journal={IEEE Transactions on Antennas and Propagation},
  volume={69},
  number={9},
  pages={5931--5941},
  year={2021},
  publisher={IEEE}
}

@article{simons2019embedding,
  title={Embedding a Rydberg atom-based sensor into an antenna for phase and amplitude detection of radio-frequency fields and modulated signals},
  author={Simons, Matthew T and Haddab, Abdulaziz H and Gordon, Joshua A and Novotny, David and Holloway, Christopher L},
  journal={IEEE Access},
  volume={7},
  pages={164975--164985},
  year={2019},
  publisher={IEEE}
}

@article{song2017quantum,
  title={Quantum-based determination of antenna finite range gain by using Rydberg atoms},
  author={Song, Zhenfei and Feng, Zhigang and Liu, Xinmeng and Li, Dabo and Zhang, Hao and Liu, Jiasheng and Zhang, Linjie},
  journal={IEEE Antennas and Wireless Propagation Letters},
  volume={16},
  pages={1589--1592},
  year={2017},
  publisher={IEEE}
}

@article{zhang2024image,
  title={Image Transmission Utilizing Amplitude Modulation in Rydberg Atomic Antenna},
  author={Zhang, Peng and Yuan, Shaoxin and Jing, Mingyong and Yuan, Jinpeng and Zhang, Hao and Zhang, Linjie},
  journal={IEEE Photonics Journal},
  year={2024},
  volume={16},
  number={2},
  pages={1-7},
  doi={10.1109/JPHOT.2024.3372640}}

@article{mao2023digital,
  title={Digital beamforming and receiving array research based on Rydberg field probes},
  author={Mao, Ruiqi and Lin, Yi and Fu, Yunqi and Ma, Yuemin and Yang, Kai},
  journal={IEEE Transactions on Antennas and Propagation},
  year={2023},
  volume={72},
  number={2},
  pages={2025-2029},
  publisher={IEEE}
}

@inproceedings{shi2023new,
  title={A New Antenna Near-Field Measurement Method based on Rydberg Atoms},
  author={Shi, Yuansheng and Ren, Wu and Li, Weiming and Cao, Meng and Lei, Mingwei and Xue, Zhenghui and Shi, Meng},
  booktitle={2023 International Conference on Microwave and Millimeter Wave Technology (ICMMT)},
  pages={1--3},
  year={2023},
  organization={IEEE}
}

@article{simons2018electromagnetically,
  title={Electromagnetically Induced Transparency (EIT) and Autler-Townes (AT) splitting in the presence of band-limited white Gaussian noise},
  author={Simons, Matthew T and Kautz, Marcus D and Holloway, Christopher L and Anderson, David A and Raithel, Georg and Stack, Daniel and St John, Marc C and Su, Wansheng},
  journal={Journal of Applied Physics},
  volume={123},
  number={20},
  year={2018},
  publisher={AIP Publishing}
}

@article{schlossberger2024rydberg,
  title={Rydberg states of alkali atoms in atomic vapour as SI-traceable field probes and communications receivers},
  author={Schlossberger, Noah and Prajapati, Nikunjkumar and Berweger, Samuel and Rotunno, Andrew P and Artusio-Glimpse, Alexandra B and Simons, Matthew T and Sheikh, Abrar A and Norrgard, Eric B and Eckel, Stephen P and Holloway, Christopher L},
  journal={Nature Reviews Physics},
  volume={6},
  number={10},
  pages={606--620},
  year={2024},
  publisher={Nature Publishing Group UK London}
}

%}

%\begin{thebibliography}{1}


%\end{thebibliography}


%\end{IEEEbiography}
%% insert where needed to balance the two columns on the last page with
%%\newpage


%\vfill


\end{document}